\documentclass{article}

\usepackage[preprint]{neurips_2025}

\usepackage[dvipsnames]{xcolor}         
\usepackage{wrapfig}
\usepackage[utf8]{inputenc} 
\usepackage[T1]{fontenc}    
\usepackage{hyperref}       
\usepackage{url}            
\usepackage{booktabs}       
\usepackage{amsfonts}       
\usepackage{nicefrac}       
\usepackage{microtype}      
\usepackage{pifont}
\usepackage{makecell}
\usepackage{tikz}

\usepackage{booktabs}
\usepackage{subfigure}
\usepackage{enumitem}
\usepackage{rotating}
\usepackage{amsmath,amssymb,amsfonts}
\usepackage[most]{tcolorbox}
\usepackage{listings}      
\usepackage{xcolor}        
\usepackage{multirow}
\usepackage{booktabs,bm}
\usepackage{algorithm}
\usepackage{algorithmic}
\usepackage{colortbl}
\definecolor{grey}{rgb}{0.89,0.71,0.57}
\definecolor{pink}{rgb}{1,0.94,1}
\definecolor{purple}{rgb}{0.84,0.78,1}
\definecolor{white}{rgb}{1,1,1}
\definecolor{backred}{RGB}{255, 190, 190}
\definecolor{backblue}{RGB}{210, 230, 250}
\definecolor{mygrey}{RGB}{200,200,200}

\hypersetup{
    colorlinks=true,
    linkcolor=red,
    citecolor=cyan,
    filecolor=magenta,      
    urlcolor=magenta,
    }

\definecolor{codegreen}{rgb}{0,0.6,0}
\definecolor{codegray}{rgb}{0.5,0.5,0.5}
\definecolor{codepurple}{rgb}{0.58,0,0.82}
\definecolor{backcolour}{rgb}{0.95,0.95,0.92}

\lstdefinestyle{mystyle}{
    backgroundcolor=\color{backcolour},   
    commentstyle=\color{codegreen},
    keywordstyle=\color{magenta},
    numberstyle=\tiny\color{codegray},
    stringstyle=\color{codepurple},
    basicstyle=\footnotesize,
    breakatwhitespace=false,         
    breaklines=true,                 
    captionpos=b,                    
    keepspaces=true,                 
    numbers=left,                    
    numbersep=5pt,                  
    showspaces=false,                
    showstringspaces=false,
    showtabs=false,                  
    tabsize=2
}

\usepackage{xspace}
\newcommand{\llmname}[1]{{\fontfamily{pcr}\selectfont {#1}}\xspace}

\makeatletter
\renewcommand{\@fnsymbol}[1]{
  \ensuremath{%
    \ifcase#1%
      \or \dagger\or \ddagger\or \mathsection\or \mathparagraph\or \|\or **\or \dagger\dagger\or \ddagger\ddagger%
    \fi
  }%
}
\makeatother

\title{ 
\method{}: Do Large Language Models Really Understand Cell Biology?
}
\def\method{\textsc{CellVerse}}

\author{
Fan Zhang$^{1}$,~~Tianyu Liu$^{2}$,~~Zhihong Zhu$^3$,~~Hao Wu$^{4}$,~~Haixin Wang$^5$,\\~~\textbf{Donghao Zhou}$^1$\textbf{,}~~\textbf{Yefeng Zheng}$^6$\textbf{,}~~\textbf{Kun Wang}$^{7,}$\thanks{Corresponding authors.}~~\textbf{,}~~\textbf{Xian Wu}$^{8,\dagger}$\textbf{,}~~\textbf{Pheng-Ann Heng}$^{1,\dagger}$\\
$^1$CUHK, $^2$Yale University, $^3$Peking University, $^4$Tsinghua University,\\ $^5$UCLA, $^6$Westlake University, $^7$NTU, $^8$Tencent\\
\texttt{fanzhang.karl@gmail.com, pheng@cse.cuhk.edu.hk}\\
Project Page: \url{https://cellverse-cuhk.github.io}
}

\begin{document}

\maketitle

\begin{abstract}

Recent studies have demonstrated the feasibility of modeling single-cell data as natural languages and the potential of leveraging powerful large language models (LLMs) for understanding cell biology. 
However, a comprehensive evaluation of LLMs' performance on language-driven single-cell analysis tasks remains unexplored.
Motivated by this challenge, we introduce \method{}, a unified language-centric question-answering benchmark that integrates four types of single-cell multi-omics data and encompasses three hierarchical levels of single-cell analysis tasks: cell type annotation (cell-level), drug response prediction (drug-level), and perturbation analysis (gene-level).
Going beyond this, we systematically evaluate the performance across 14 open-source and closed-source LLMs ranging 160M $\rightarrow$ 671B on \method{}. Remarkably, the experimental results reveal:
\ding{172} Existing specialist models (\emph{e.g.}, \llmname{C2S-Pythia}) fail to make reasonable decisions across all sub-tasks within \method{}, while generalist models such as \llmname{Qwen}, \llmname{Llama}, \llmname{GPT}, and \llmname{DeepSeek} family models exhibit preliminary understanding capabilities within the realm of cell biology.
\ding{173} The performance of current LLMs falls short of expectations and has substantial room for improvement.
Notably, in the widely studied drug response prediction task, \underline{none} of the evaluated LLMs demonstrate significant performance improvement over \textit{random guessing}.
\method{} offers the first large-scale empirical demonstration that significant challenges still remain in applying LLMs to cell biology. 
By introducing \method{}, we lay the foundation for advancing cell biology through natural languages and hope this paradigm could facilitate next-generation single-cell analysis.

\end{abstract}

\section{Introduction}\label{sec:intro}

\begin{figure}[t]
    \centering
    \includegraphics[width=\linewidth]{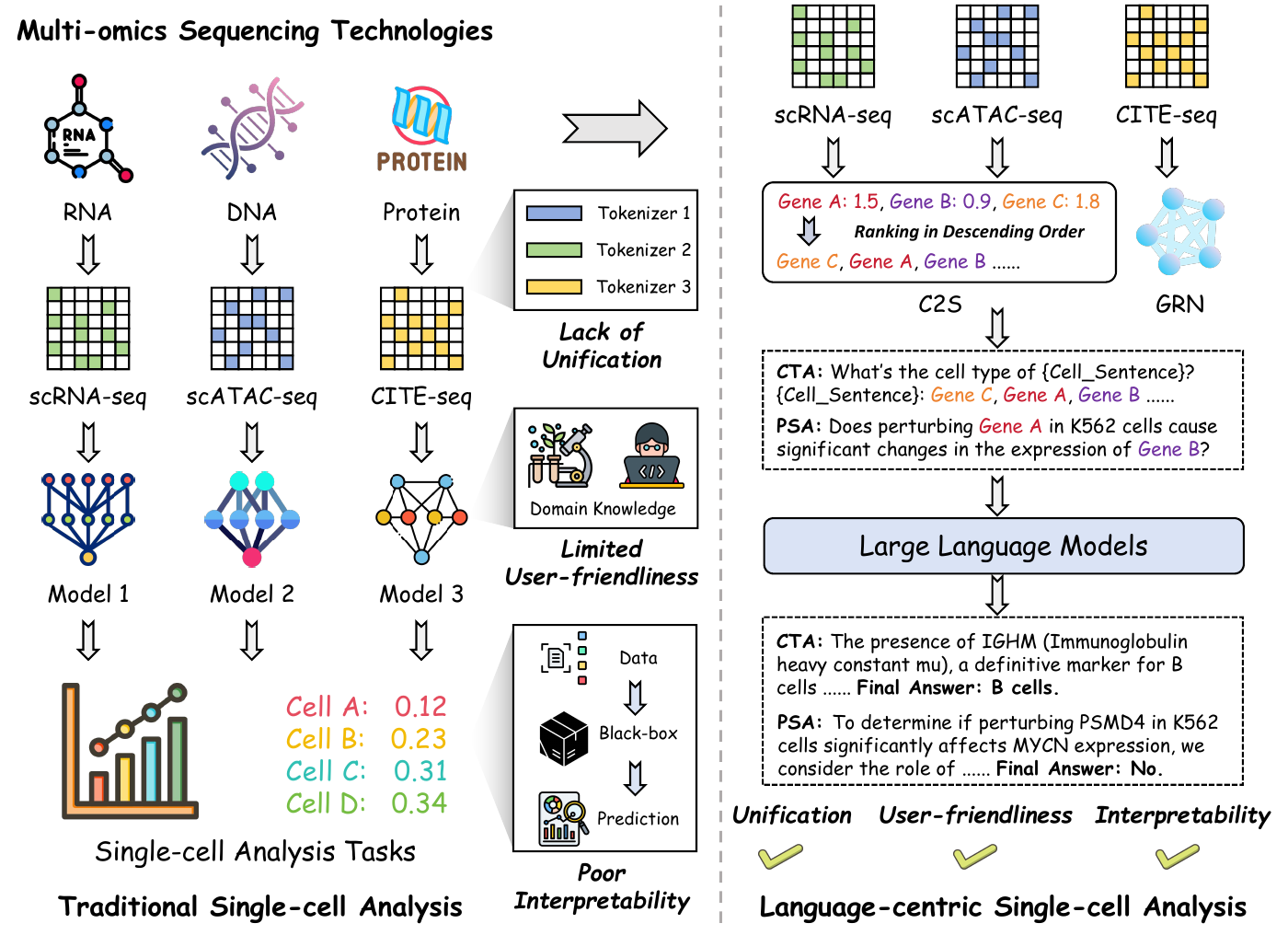}
    \vspace{-4mm}
    \caption{An illustration of traditional single-cell analysis and language-centric single-cell analysis.}
    \vspace{-3mm}
    \label{fig:illus}
\end{figure}

Single-cell analysis \cite{lindstrom2010overview,lee2020single,heumos2023best} has received growing attention in recent years due to its powerful capabilities across a wide range of healthcare applications, including disease diagnosis \cite{brandow2022advances}, drug discovery \cite{heath2016single}, and immunotherapy \cite{june2018car}. 
With the rapid progress in artificial intelligence and deep learning \cite{lecun2015deep,vaswani2017attention}, methodological advances in single-cell analysis have undergone a notable shift from traditional statistical techniques \cite{korthauer2016statistical,argelaguet2020mofa+} to specialized deep learning models \cite{li2020deep,ma2022deep}, and more recently, to approaches based on large-scale pre-trained foundation models \cite{cui2024scgpt,hao2024large}. 
While these developments have led to significant improvements in performance and scalability, several inherent limitations remain unresolved (Figure \ref{fig:illus}):
(1) \textit{Lack of Unification.} For different types of omics data and downstream tasks, existing paradigms typically require separately designed models, lacking a unified approach capable of simultaneously handling multi-omics and multi-task scenarios.
(2) \textit{Limited User-Friendliness.} Effective application of these methods to single-cell analysis often necessitates domain expertise in biology as well as proficiency in programming. Furthermore, the lack of user-centric interaction design in current models poses a significant barrier to adoption by non-expert users.
(3) \textit{Poor Interpretability.} Most of the existing data-driven black-box models directly learn the mapping from input (\emph{e.g.}, gene expression) to output (\emph{e.g.}, cell type information), without incorporating interpretable intermediate steps. As a result, users are often unable to understand the rationale behind the model's decisions.
To this end, we seek to establish a unified, user-friendly, and interpretable paradigm for single-cell analysis. Leveraging the advanced techniques, such as cell2sentence \cite{levine2024cell2sentence} and gene regulatory network \cite{karlebach2008modelling,davidson2005gene}, we can convert single-cell data into natural language formats.
This transformation allows us to reformulate single-cell analysis tasks as question-answering (QA) problems—an interface that aligns with the capabilities of large language models (LLMs) and facilitates end-to-end reasoning over domain-specific knowledge in cell biology.
With the reasoning capabilities and easy deployment of advanced LLMs, even non-expert users can efficiently analyze single-cell data, with interpretability supported via explicit querying of reasoning trajectories.

Specifically, we introduce \method{}, a unified language-centric benchmark dataset for evaluating the capabilities of LLMs in single-cell analysis. We begin by curating five sub-datasets spanning four types of single-cell multi-omics data (scRNA-seq \cite{schirmer2019neuronal}, CITE-seq \cite{mimitou2021scalable}, ASAP-seq \cite{mimitou2021scalable}, and scATAC-seq data \cite{adamson2016multiplexed}) and translate them into natural languages. Subsequently, we select three most representative single-cell analysis tasks—cell type annotation \cite{pasquini2021automated} (cell-level), drug response prediction \cite{adam2020machine} (drug-level), and perturbation analysis \cite{ji2021machine} (gene-level)—and reformulate them as QA problems by integrating each with the natural language-formatted single-cell data.
Next, we conduct a comprehensive and systematic evaluation of 14 advanced LLMs on the proposed \method{} benchmark. The evaluated models include open-source LLMs such as \llmname{C2S-Pythia (160M, 410M, and 1B)} \cite{levine2024cell2sentence,rizvi2025scaling}, \llmname{Qwen-2.5 (7B, 32B, and 72B)} \cite{yang2024qwen2}, \llmname{Llama-3.3–70B} \cite{grattafiori2024llama}, and \llmname{DeepSeek (V3 and R1)} \cite{liu2024deepseek,guo2025deepseek}, as well as closed-source models including \llmname{GPT-4} \cite{achiam2023gpt}, \llmname{GPT-4o-mini} \cite{openaigpt4omini}, \llmname{GPT-4o} \cite{openaigpt4o}, \llmname{GPT-4.1-mini} \cite{openaigpt41}, and \llmname{GPT-4.1} \cite{openaigpt41}. 

Through a comprehensive analysis of the experimental results, we observe the following key findings:
(1) Specialist models (\llmname{C2S-Pythia}), despite being trained specifically on single-cell analysis tasks, consistently exhibit hallucination issues across all tasks. Due to limited model capacity and insufficient training data, they fail to make accurate decisions. In contrast, generalist models, though not fine-tuned for single-cell analysis, display initial reasoning capabilities and perform reasonably across various tasks. More importantly, some models not only produce task-specific predictions but also generate complete reasoning paths \cite{wei2022chain}, demonstrating the potential of advanced LLMs for understanding cell biology.
(2) However, generalist models still fall short of expectations. For the cell type annotation task, the state-of-the-art (SOTA) model achieves accuracies of $42.38\%$ on scRNA-seq data, $61.43\%$ on CITE-seq data, and $29.33\%$ on ASAP-seq data, indicating considerable room for improvement. On the more challenging tasks of drug response prediction and perturbation analysis (significance and direction), SOTA accuracies reach $55\%$, $76.67\%$, and $62.96\%$, respectively—most of them do not significantly outperform random guessing.

In summary, the main contributions of this paper are as follows:
\begin{itemize}[leftmargin=*]
    \item We identify key limitations in unification, user-friendliness, and interpretability of existing single-cell analysis paradigms. To address these issues, we propose a novel perspective: transforming single-cell data into natural languages and leveraging advanced LLMs for language-driven analysis. 
    \item We propose \method{}, a unified language-centric benchmark dataset for single-cell analysis that covers four types of single-cell multi-omics data and three representative sub-tasks. To the best of our knowledge, \method{} is the first dataset designed to evaluate the understanding capabilities of LLMs in the domain of cell biology, and serves as a foundation for future research in this area.
    \item We conduct a comprehensive and systematic evaluation of 14 open-source and closed-source advanced LLMs on \method{}, accompanied by in-depth analysis. Our analyses and findings offer insights and potential directions for future research in applying LLMs to cell biology.
\end{itemize}

\section{Related Work}

\subsection{Large Language Models for Scientific Problems}
Recent advances in large language models (LLMs) have spurred interest in applying them to scientific domains, including mathematics \cite{imani2023mathprompter,ahn2024large}, chemistry \cite{pan2023large,zheng2025large}, and biology \cite{luo2022biogpt,madani2023large}.
For example, SciBERT \cite{beltagy2019scibert} and BioBERT \cite{lee2020biobert} leverage domain-specific pretraining on biomedical corpora to enhance their performance in biomedical text mining tasks, such as named entity recognition and relation extraction. Galactica \cite{taylor2022galactica} strives to unify scientific knowledge representation and generation, whereas AlphaCode \cite{li2022competition} expands LLMs into program synthesis for scientific computation. 
These models exhibit promising early achievements in various tasks, including summarizing scientific literature, symbolic reasoning, and tackling textbook-style problems.
However, most applications focus on general-purpose scientific texts \cite{cai2024sciassess} or structured symbolic inputs \cite{xu2023symbol}. In contrast, the utilization of LLMs for domain-specific and data-intensive tasks, such as single-cell analysis, remains relatively unexplored. Our work contributes to this emerging field by connecting LLMs with scientific biological data, aiming to evaluate and improve LLM capabilities in real-world biological contexts.

\subsection{Benchmark Datasets for Science QA}
A variety of question-answering (QA) benchmark datasets have been proposed to evaluate the performance of advanced LLMs in scientific domains \cite{sun2024scieval,chen2021geoqa}. 
Notable examples include SciBench \cite{wang2023scibench}, MathVista \cite{lu2023mathvista}, and PubMedQA \cite{jin2019pubmedqa}, which cover college-level scientific problems, multimodal mathematical reasoning problems, and biomedical literature, respectively. 
These benchmarks primarily focus on evaluating the capability of textual understanding and reasoning with easy-to-handle scientific contexts.
However, none of the existing benchmarks are applicable to cell biology or account for the high-dimensional and sparse nature of single-cell data.
Our work fills this gap by introducing a benchmark tailored to evaluating LLMs on biological QA tasks at the single-cell level.

\begin{figure}[t]
    \centering
    \includegraphics[width=\linewidth]{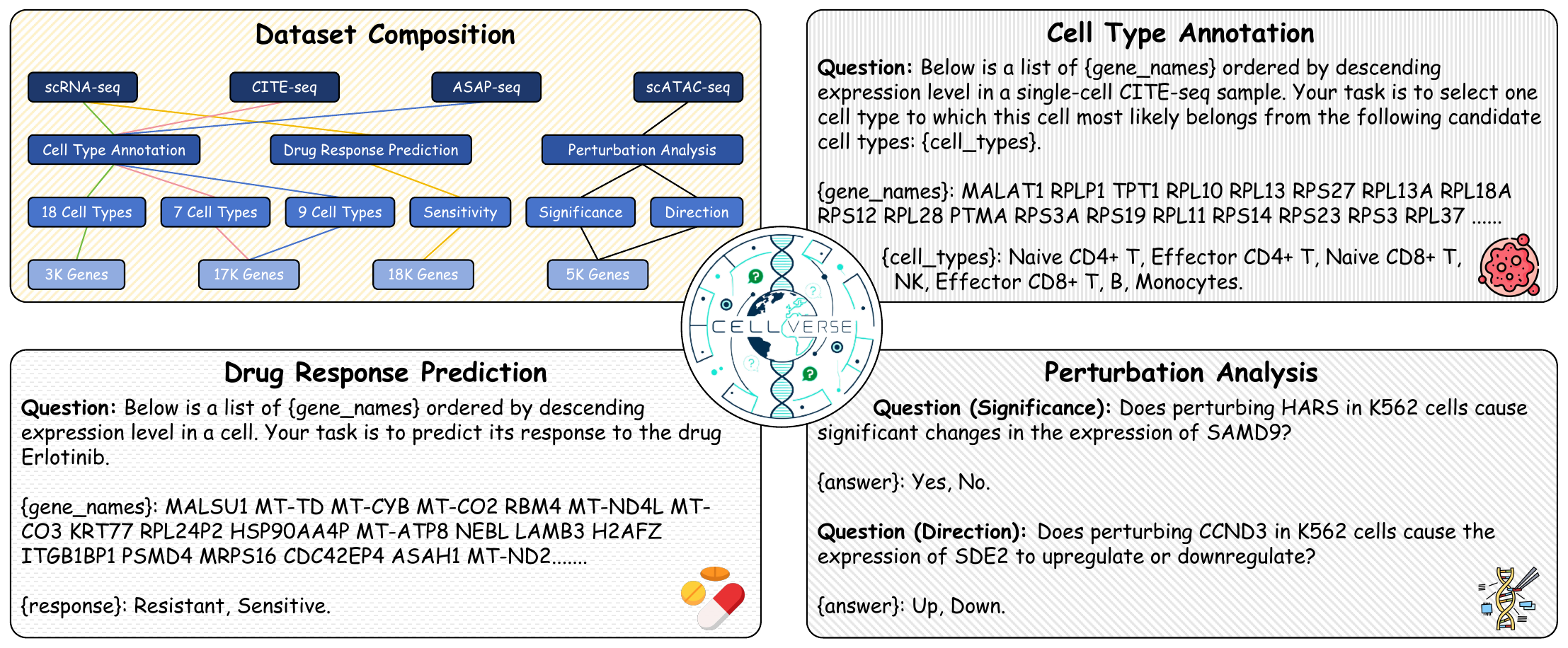}
    \vspace{-3mm}
    \caption{An overview of \method{}. The top-left panel illustrates the composition of \method{}, while the remaining three panels provide concrete data examples within \method{}.}
    \label{fig:cellverse}
\end{figure}
\section{The \method{} Dataset}
In this section, we will introduce the proposed \method{} dataset. Firstly, in Section \ref{sec:pre_background}, we provide relevant preliminaries and background information. Then, in Section \ref{sec:data_curation}, we detail the dataset curation process. Finally, in Section \ref{sec:data_stat}, we present some key statistics of the dataset.

\subsection{Preliminaries and Background}\label{sec:pre_background}
Single-cell data are characterized by long sequences \cite{picelli2014full}, high sparsity \cite{baruzzo2020sparsim}, substantial noise \cite{brennecke2013accounting}, and strong heterogeneity \cite{buettner2015computational}. Given a single-cell dataset $X \in \mathbb{R}^{N \times G}$, where $N$ denotes the number of cells and $G$ represents the number of features (typically corresponding to genes), the data structure varies across sequencing technologies and omics types. Depending on the modality, single-cell data may capture gene expression (scRNA-seq data), chromatin accessibility (scATAC-seq data), protein abundance (CITE-seq data), or DNA peak signals (ASAP-seq data).
Due to the inherent input format and context length limitations of LLMs, efficiently transforming raw single-cell data into natural languages that retain key information with minimal loss remains a significant challenge. In the following sections, we introduce two feasible approaches to address this problem: (1) \textit{cell2sentence (C2S)}, which encodes cell-level information into natural language, and (2) \textit{gene regulatory network (GRN)}, which converts gene-level interactions into interpretable texts.

\paragraph{Cell2sentence.} 
To transform single-cell data into natural languages, cell2sentence (C2S) \cite{levine2024cell2sentence} offers an intuitive perspective: it connects normalized expression profiles to natural languages by leveraging gene names and their ranked expression levels. Specifically, C2S treats gene names as tokens and represents each cell as a sentence composed of the top $n$ most highly expressed genes in descending order. This process can be formulated as:
\begin{equation}
    \text{Cell Sentence}_i = \texttt{[}g_{i}^{(1)}, g_{i}^{(2)}, \dots, g_{i}^{(n)}\texttt{]},
\end{equation}

where $g_{i}^{(j)}$ denotes the gene name ranked $j$-th in expression level for cell $x_i \in \mathbb{R}^{1 \times G}$. This translation preserves essential cellular characteristics in a compact and interpretable form with controllable context length, facilitating the application of LLMs to high-dimensional single-cell data.

\paragraph{Gene Regulatory Network.}
To further enable natural language understanding of cell biology at the gene level, we leverage the gene regulatory network (GRN) \cite{karlebach2008modelling,davidson2005gene}, which offers structured information of gene–gene interactions. A GRN is typically modeled as a directed graph $\mathcal{G} = (\mathcal{V}, \mathcal{E})$, where nodes $\mathcal{V}$ represent genes and edges $\mathcal{E}$ represent regulatory relationships inferred from expression data or external databases. 
Each edge $ (g^a, g^b) \in \mathcal{E} $ with an associated weight is then translated into a natural language statement. Specifically, a nonzero edge weight implies that perturbing gene $g^a$ leads to a change in the expression of gene $g^b$. Conversely, if no edge exists or the weight is negligible, it suggests that perturbing $g^a$ does not significantly affect $g^b$.
This process can be formulated as:
\begin{equation}
    (g^a, g^b, w_{ab}) \in \mathcal{E}' \quad \Rightarrow \quad 
\begin{cases}
\delta (g^a) \rightarrow \text{Change} (g^b) , & \text{if } w_{ab} \geq \tau \\
\delta (g^a) \nrightarrow \text{Change} (g^b), & \text{if } w_{ab} < \tau
\end{cases},
\end{equation}
where $ w_{ab} $ is the edge weight indicating the regulatory strength from $g^a$ to $g^b$, $\delta$ is the perturbation operator, and $ \tau $ is a predefined threshold to determine biological significance.
This transformation offers an interpretable summary of gene regulatory mechanisms, allowing LLMs to reason about biological pathways and perturbations. By expressing these regulatory dependencies in natural language, we effectively bridge graph-based biological knowledge and language-driven inference.

\subsection{Data Curation}\label{sec:data_curation}
With the help of C2S and GRN, we can transform various single-cell analysis tasks into a unified QA format. Next, we detail the data curation process for \method{}, which encompasses three core tasks: cell type annotation, drug response prediction, and perturbation analysis.

\paragraph{Cell Type Annotation.}
For the cell type annotation task, we first use C2S to convert raw single-cell data into cell sentences, and then directly query the corresponding cell type for each sentence. After generating the initial QA pairs, we perform two post-processing steps. First, we filter out low-quality samples with excessive redundancy based on sentence similarity. Second, we apply a resampling strategy to promote a more balanced data distribution across all cell types. An example of the resulting data for the cell type annotation task is shown in the top-right part of Figure \ref{fig:cellverse}.

\paragraph{Drug Response Prediction.}
For the drug response prediction task, we similarly use C2S to convert single-cell data into cell sentences, then query each sentence's response to a specific drug. After generating the QA pairs, we perform the same two post-processing steps as in the previous task: filtering out low-quality samples with high redundancy and promoting a more balanced data distribution. An example for drug response prediction is shown in the bottom-left panel of Figure \ref{fig:cellverse}.

\paragraph{Perturbation Analysis.}
For the perturbation analysis task, we leverage the GRN to identify interactions between genes. Specifically, for each candidate gene pair, we perform a non-parametric Wilcoxon test \cite{gehan1965generalized,cuzick1985wilcoxon} between the perturbed and control groups. An interaction is considered significant if the p-value is below 0.05 and the log$_2$ fold change (log$_2$FC) exceeds 0.5, indicating that perturbing gene $g^a$ induces a statistically significant change in the expression of gene $g^b$.
Based on this criterion, we define two sub-tasks: (1) perturbation significance analysis, which asks whether perturbing gene $g^a$ significantly affects gene $g^b$; and (2) perturbation direction analysis, which further queries whether the expression of gene $g^b$ increases or decreases following the perturbation of gene $g^a$.
To ensure data quality and coverage, we retain only samples with more than 10 cells in both perturbed and control groups and limit each source gene to a maximum of three QA examples.
An example for the above two tasks is presented in the bottom-right panel of Figure~\ref{fig:cellverse}.

\subsection{Data Statistics}\label{sec:data_stat}
\begin{wrapfigure}{r}{0.55\textwidth}\vspace{-5mm}
    \centering
    \includegraphics[width=\linewidth]{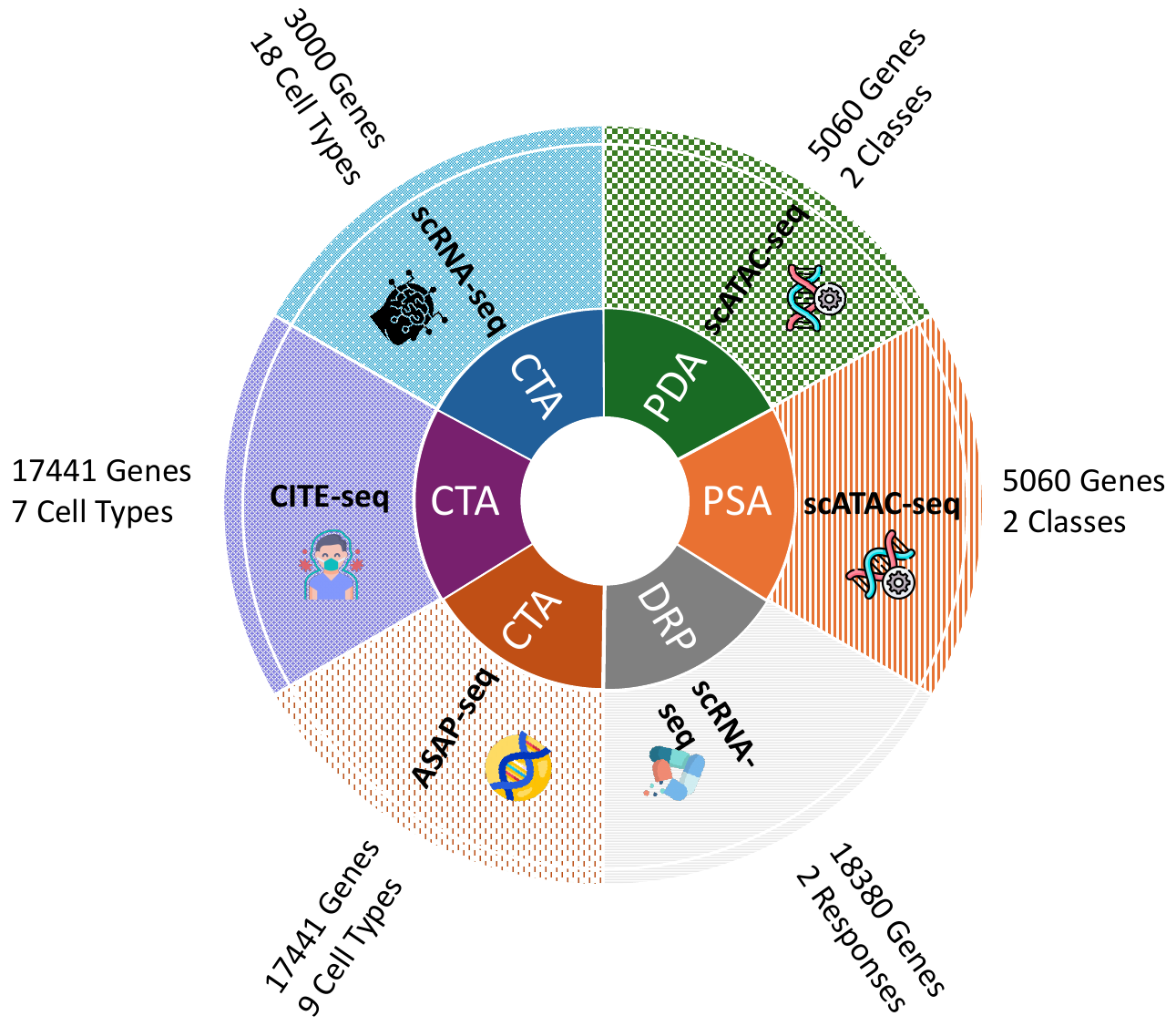}
    \vspace{-3mm}
    \caption{Data statistics information of \method{}.}
    \vspace{-3mm}
    \label{fig: data_stat}
\end{wrapfigure}
After completing the data curation process, we construct the \method{} dataset, which integrates four types of single-cell multi-omics data and spans three sub-tasks (as illustrated in Figure \ref{fig: data_stat}).
\textbf{Cell Type Annotation (CTA).} We include data from three different omics modalities:
(1) scRNA-seq Multiple Sclerosis data \cite{schirmer2019neuronal}: This dataset contains 3{,}000 gene expression profiles and 18 annotated cell types: \textit{phagocyte}, \textit{cortical layer 2–3 excitatory neuron A}, \textit{cortical layer 4 excitatory neuron}, \textit{mixed glial cell}, \textit{SV2C-expressing interneuron}, \textit{microglial cell}, \textit{cortical layer 5–6 excitatory neuron}, \textit{oligodendrocyte A}, \textit{cortical layer 2–3 excitatory neuron B}, \textit{mixed excitatory neuron}, \textit{endothelial cell}, \textit{VIP-expressing interneuron}, \textit{PVALB-expressing interneuron}, \textit{oligodendrocyte precursor cell}, \textit{pyramidal neuron}, \textit{SST-expressing interneuron}, \textit{astrocyte}, and \textit{oligodendrocyte C}.
(2) CITE-seq PBMC data \cite{mimitou2021scalable}: This dataset includes 17{,}441 genes, annotated with 7 cell types: \textit{Naive CD4+ T}, \textit{Effector CD4+ T}, \textit{Naive CD8+ T}, \textit{NK}, \textit{Effector CD8+ T}, \textit{B}, and \textit{Monocytes}.
(3) ASAP-seq PBMC data \cite{mimitou2021scalable}: This dataset also includes 17{,}441 genes and 9 cell types: \textit{DC}, \textit{Naive CD4+ T}, \textit{Effector CD4+ T}, \textit{Naive CD8+ T}, \textit{NK}, \textit{Effector CD8+ T}, \textit{B}, \textit{Monocytes}, and \textit{unknown}.
\textbf{Drug Response Prediction (DRP).} We utilize a scRNA-seq dataset \cite{aissa2021single} measuring cellular responses to the drug erlotinib. This dataset contains 18{,}380 genes and two drug responses: \textit{sensitive} and \textit{resistant}.
\textbf{Perturbation Analysis (PA).} For this task, we utilize scATAC-seq data from K562 cells \cite{adamson2016multiplexed}, encompassing 5{,}060 genes, to define two sub-tasks:
\textbf{Perturbation Significance Analysis (PSA).} A binary classification task that asks whether perturbing gene \( g^a \) significantly affects gene \( g^b \), with answer choices \textit{yes} or \textit{no}.
\textbf{Perturbation Direction Analysis (PDA).} Another binary classification task that determines the direction of change for gene \( g^b \) after perturbation of gene \( g^a \), with answer choices \textit{up} or \textit{down}.
\vspace{-3mm}

\section{Experiments}

\begin{figure}[t]
    \centering
    \includegraphics[width=\linewidth]{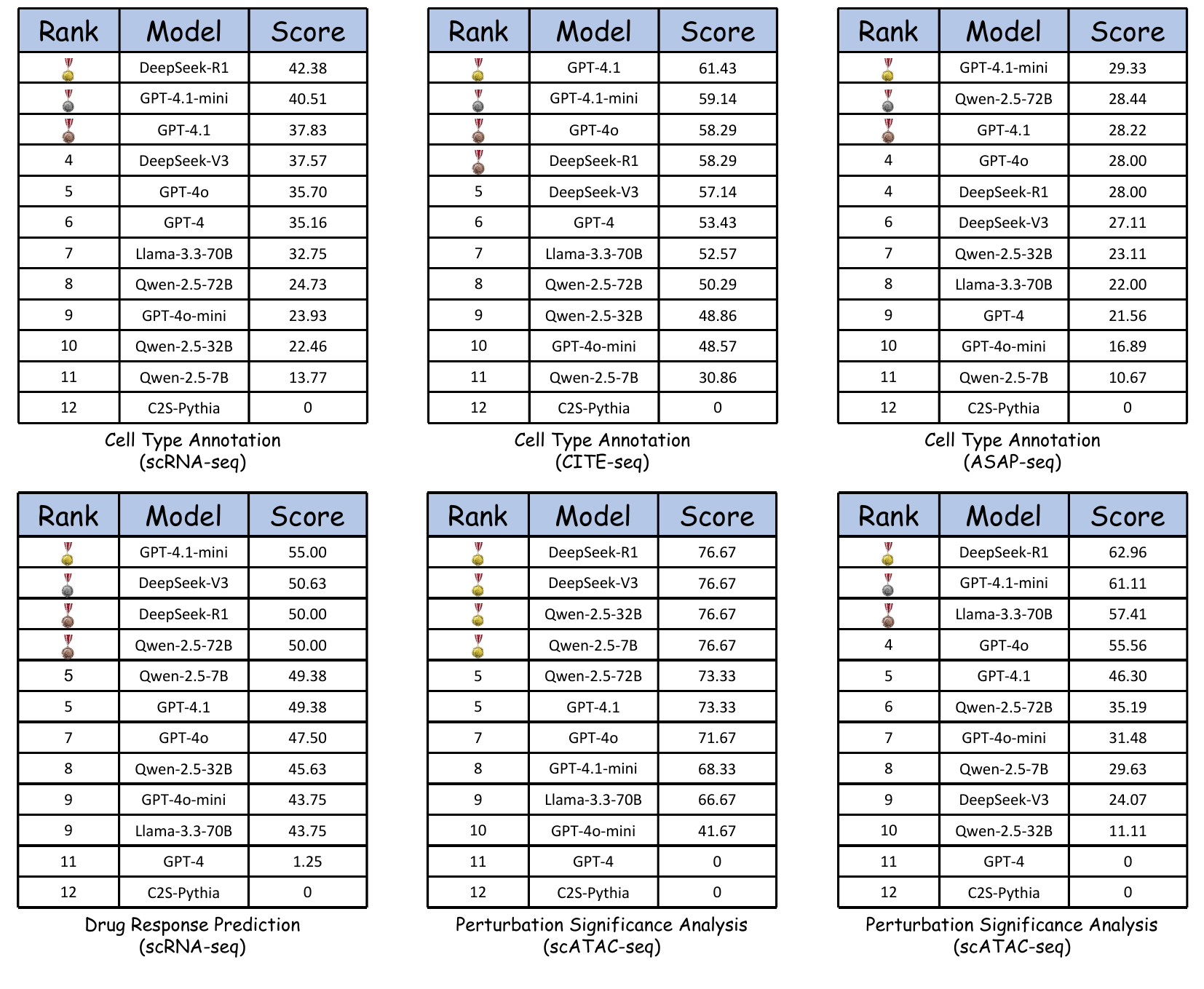}
    \vspace{-5mm}
    \caption{Leaderboard results on our \method{} benchmark. The scores represent the prediction accuracy of LLMs. We include both open-source and closed-source LLMs in the evaluation.}
    \label{fig:leaderboard}
    \vspace{-4mm}
\end{figure}

\subsection{Evaluation Protocols and Implementation Details}\label{sec:protocol_implement}
To evaluate the performance of current advanced LLMs on the proposed \method{}, we conduct a fair comparison across 9 open-source models (\llmname{C2S-Pythia-160M} \cite{levine2024cell2sentence}, \llmname{C2S-Pythia-410M} \cite{levine2024cell2sentence}, \llmname{C2S-Pythia-1B} \cite{rizvi2025scaling}, \llmname{Qwen-2.5-7B} \cite{yang2024qwen2}, \llmname{Qwen-2.5-32B} \cite{yang2024qwen2}, \llmname{Qwen-2.5-72B} \cite{yang2024qwen2}, \llmname{Llama-3.3–70B} \cite{grattafiori2024llama}, \llmname{DeepSeek-V3} \cite{liu2024deepseek}, and \llmname{DeepSeek-R1} \cite{guo2025deepseek}) and 5 closed-source models (\llmname{GPT-4} \cite{achiam2023gpt}, \llmname{GPT-4o-mini} \cite{openaigpt4omini}, \llmname{GPT-4o} \cite{openaigpt4o}, \llmname{GPT-4.1-mini} \cite{openaigpt41}, and \llmname{GPT-4.1} \cite{openaigpt41}).
Among these, only the \llmname{C2S-Pythia} series models were trained specifically on single-cell analysis tasks and thus considered specialist models, while all others are generalist models.
We perform inference using the vLLM framework \cite{kwon2023efficient} for open-source models and official APIs for closed-source models.
During preliminary testing, we observed that all models struggle to make reasonable decisions under open-ended question settings. Therefore, we convert all questions into multiple-choice format by including a list of candidate answers directly in the prompt. 
The experiments are conducted under both zero-shot and few-shot settings.
Model performance is then evaluated using standard metrics: precision score, recall score, F1 score, and overall accuracy.

\begin{table}[t]
\centering
\caption{Performance comparison ($\%$) of cell type annotation on single-cell multi-omics data.
}
\label{tab:cta}
\resizebox{\columnwidth}{!}{%
\begin{tabular}{l|cccc|cccc|cccc}
\toprule
\textbf{Data} & \multicolumn{4}{c|}{\textbf{scRNA-seq}} & \multicolumn{4}{c|}{\textbf{CITE-seq}} & \multicolumn{4}{c}{\textbf{ASAP-seq}} \\
\midrule
\textbf{Metric} & \textbf{Precision} & \textbf{Recall} & \textbf{F1} & \textbf{Accuracy} & \textbf{Precision} & \textbf{Recall} & \textbf{F1} & \textbf{Accuracy} & \textbf{Precision} & \textbf{Recall} & \textbf{F1} & \textbf{Accuracy} \\
\midrule
\rowcolor{black!10}\multicolumn{13}{c}{\textit{Open-source LLMs}} \\
\midrule
\llmname{Qwen-2.5-7B} & 36.31 & 14.54 & 11.53 & 13.77 & 41.42 & 27.00 & 28.08 & 30.86 & 15.47 & 9.60 & 7.37 & 10.67 \\
\llmname{Qwen-2.5-32B} & 28.81 & 20.60 & 18.76 & 22.46 & 45.58 & 42.75 & 38.57 & 48.86 & 25.87 & 20.80 & 19.87 & 23.11 \\
\llmname{Qwen-2.5-72B} &30.94&22.39& 19.32 &24.73& 43.60& 44.00& 40.50& 50.29& 25.04& 25.60& 22.61& \colorbox{backblue!75}{28.44}  \\
\llmname{Llama-3.3-70B} &33.60& 30.25&24.99& 32.75& 61.30& 52.57& 50.02& 52.57& 26.95& 22.00 &19.30 &22.00 \\
\llmname{DeepSeek-V3} & 38.69 & 34.99 & 30.60 & 37.57 &  \colorbox{backblue!75}{66.44} & 57.14 & 54.65 & 57.14 & 26.52 & 24.40 & 21.18 & 27.11 \\
\llmname{DeepSeek-R1} & 39.95 & \colorbox{backred!50}{38.81} & \colorbox{backblue!75}{33.40} & \colorbox{backred!50}{42.38} & 57.57 & 51.00 & 50.76 & 58.29 & 32.69 & 25.20 & 20.81 & 28.00\\
\midrule
\rowcolor{black!10}\multicolumn{13}{c}{\textit{Closed-source LLMs}} \\
\midrule
\llmname{GPT-4} & 40.81 & 32.41 & 29.89 & 35.16 & 47.21 & 46.75 & 43.44 & 53.43 & 26.48 & 19.40 & 17.16 & 21.56 \\
\llmname{GPT-4o-mini} & 31.86 & 22.64 & 15.77 & 23.93 & 63.94 & 48.57 & 47.89 & 48.57 & 33.27 & 16.89 & 13.52 & 16.89 \\
\llmname{GPT-4o} & 40.24 & 31.52 & 29.03 & 35.70 & 60.07 & 58.29 & 55.76 & 58.29 & 32.47 & 28.00 & \colorbox{backblue!75}{25.00} & 28.00 \\
\llmname{GPT-4.1-mini} & \colorbox{backblue!75}{41.14} & \colorbox{backblue!75}{36.46} & \colorbox{backred!50}{34.55} & \colorbox{backblue!75}{40.51} & 66.13 & \colorbox{backblue!75}{59.14} & \colorbox{backblue!75}{58.47} & \colorbox{backblue!75}{59.14} & \colorbox{backblue!75}{36.86} & \colorbox{backred!50}{29.33} & \colorbox{backred!50}{27.47} & \colorbox{backred!50}{29.33} \\
\llmname{GPT-4.1} & \colorbox{backred!50}{42.46} & 35.19 & 30.94 & 37.83 & \colorbox{backred!50}{68.32} & \colorbox{backred!50}{61.43} & \colorbox{backred!50}{59.26} & \colorbox{backred!50}{61.43} & \colorbox{backred!50}{42.53} & \colorbox{backblue!75}{28.22} & 23.99 & 28.22 \\
\bottomrule
\end{tabular}%
}
\end{table}

\begin{table}[t]
\centering
\caption{Performance comparison ($\%$) of drug response prediction on scRNA-seq data. }
\label{tab:drp}
\resizebox{\columnwidth}{!}{%
\begin{tabular}{l|cccc|ccc|ccc}
\toprule
\textbf{Setting} & \multicolumn{4}{c|}{\textbf{Overall}} & \multicolumn{3}{c|}{\textbf{Sensitive}} & \multicolumn{3}{c}{\textbf{Resistant}} \\
\midrule
\textbf{Metric} & \textbf{Precision} & \textbf{Recall} & \textbf{F1} & \textbf{Accuracy} & \textbf{Precision} & \textbf{Recall} & \textbf{F1} & \textbf{Precision} & \textbf{Recall} & \textbf{F1} \\
\midrule
\rowcolor{black!10}\multicolumn{11}{c}{\textit{Open-source LLMs}} \\
\midrule
\llmname{Qwen-2.5-7B} & 33.82 & 32.92 & 32.38 & 49.38 & 49.52 & \colorbox{backred!50}{65.00} & \colorbox{backred!50}{56.22} & 51.92 & 33.75 & 40.91 \\
\llmname{Qwen-2.5-32B} & 44.85 & 45.63 & 43.50 & 45.63 & 42.86 & 26.25 & 32.56 & 46.85 & 65.00 & 54.45 \\
\llmname{Qwen-2.5-72B} &34.02& 33.33 &	30.84 &	50.00 &	50.00& 	21.25 &	29.82 &	\colorbox{backblue!75}{52.07} &	78.75 &	62.69 \\
\llmname{Llama-3.3-70B}& 43.71 &	43.75 &	\colorbox{backblue!75}{43.67} & 43.75 &44.19 & \colorbox{backblue!75}{47.50} & \colorbox{backblue!75}{45.78} &	43.24 &	40.00 &	41.56 \\ 
\llmname{DeepSeek-V3} & \colorbox{backblue!75}{52.44} & \colorbox{backblue!75}{50.63} & 39.34 & \colorbox{backblue!75}{50.63} & \colorbox{backblue!75}{54.55} & 7.50 & 13.19 & 50.34 & 93.75 & 65.50 \\
\llmname{DeepSeek-R1} & 25.00 & 50.00 & 33.33 & 50.00 & 0.00 & 0.00 & 0.00 & 50.00 & \colorbox{backred!50}{100.00} & \colorbox{backblue!75}{66.67} \\
\midrule
\rowcolor{black!10}\multicolumn{11}{c}{\textit{Closed-source LLMs}} \\
\midrule
\llmname{GPT-4} & 13.33 & 0.83 & 1.57 & 1.25 & 0.00 & 0.00 & 0.00 & 40.00 & 0.25 & 4.71 \\
\llmname{GPT-4o-mini} & 36.82 & 43.75 & 35.24 & 43.75 & 27.27 & 7.50 & 11.76 & 46.38 & 80.00 & 58.72 \\
\llmname{GPT-4o} & 24.78 & 31.67 & 22.96 & 47.50 & 25.00 & 2.50 & 4.55 & 49.33 & 92.50 & 64.35 \\
\llmname{GPT-4.1-mini} & \colorbox{backred!50}{60.54} & \colorbox{backred!50}{55.00} & \colorbox{backred!50}{48.19} & \colorbox{backred!50}{55.00} & \colorbox{backred!50}{68.18} & 18.75 & 29.41 & \colorbox{backred!50}{52.90} & 91.25 & \colorbox{backred!50}{66.97} \\
\llmname{GPT-4.1} & 44.84 & 49.38 & 35.12 & 49.38 & 40.00 & 2.50 & 4.71 & 49.68 & \colorbox{backblue!75}{96.25} & 65.53\\
\bottomrule
\end{tabular}%
}
\vspace{-3mm}
\end{table}

\begin{table}[t]
\centering
\caption{Performance comparison ($\%$) of perturbation significance analysis on scATAC-seq data. }
\label{tab:perturbation-significance}
\resizebox{\columnwidth}{!}{%
\begin{tabular}{l|cccc|ccc|ccc}
\toprule
\textbf{Setting} & \multicolumn{4}{c|}{\textbf{Overall}} & \multicolumn{3}{c|}{\textbf{Yes}} & \multicolumn{3}{c}{\textbf{No}} \\
\midrule
\textbf{Metric} & \textbf{Precision} & \textbf{Recall} & \textbf{F1} & \textbf{Accuracy} & \textbf{Precision} & \textbf{Recall} & \textbf{F1} & \textbf{Precision} & \textbf{Recall} & \textbf{F1} \\
\midrule
\rowcolor{black!10}\multicolumn{11}{c}{\textit{Open-source LLMs}} \\
\midrule
\llmname{Qwen-2.5-7B} & 38.33 & 50.00 & 43.40 & \colorbox{backred!50}{76.67} & 0.00 & 0.00 & 0.00 & 76.67 & \colorbox{backred!50}{100.00} & \colorbox{backred!50}{86.79} \\
\llmname{Qwen-2.5-32B} & 38.33 & 50.00 & 43.40 & \colorbox{backred!50}{76.67} & 0.00 & 0.00 & 0.00 & 76.67 & \colorbox{backred!50}{100.00} & \colorbox{backred!50}{86.79} \\
\llmname{Qwen-2.5-72B} &37.93 &47.83& 42.31 &\colorbox{backblue!75}{73.33} &0.00 &0.00 &0.00 &75.86& 95.65& 84.62  \\
\llmname{Llama-3.3-70B} &58.75 & \colorbox{backblue!75}{60.87} & \colorbox{backblue!75}{58.96} &66.67& 35.00 & \colorbox{backblue!75}{50.00} & \colorbox{backblue!75}{41.18} & \colorbox{backred!50}{82.50} &	71.74 &76.74 \\
\llmname{DeepSeek-V3} & \colorbox{backblue!75}{63.79} & 52.48 & 49.52 & \colorbox{backred!50}{76.67} & \colorbox{backred!50}{50.00} & 7.14 & 12.50 & 77.59 & \colorbox{backblue!75}{97.83} & \colorbox{backblue!75}{86.54} \\
\llmname{DeepSeek-R1} & \colorbox{backred!50}{64.81} & 57.45 & 58.00 & \colorbox{backred!50}{76.67} & \colorbox{backred!50}{50.00} & 21.43 & 30.00 & 79.63 & 93.48 & 86.00 \\
\midrule
\rowcolor{black!10}\multicolumn{11}{c}{\textit{Closed-source LLMs}} \\
\midrule
\llmname{GPT-4} & 0.00 & 0.00 & 0.00 & 0.00 & 0.00 & 0.00 & 0.00 & 0.00 & 0.00 & 0.00 \\
\llmname{GPT-4o-mini} & 53.97 & 54.50 & 41.52 & 41.67 & 25.58 & \colorbox{backred!50}{78.57} & 38.60 & \colorbox{backblue!75}{82.35} & 30.43 & 44.44 \\
\llmname{GPT-4o} & 61.11 & \colorbox{backred!50}{61.65} & \colorbox{backred!50}{61.35} & 71.67 & \colorbox{backblue!75}{40.00} & 42.86 & \colorbox{backred!50}{41.38} & 82.22 & 80.43 & 81.32 \\
\llmname{GPT-4.1-mini} & 25.79 & 29.71 & 27.61 & 68.33 & 0.00 & 0.00 & 0.00 & 77.36 & 89.13 & 82.83 \\
\llmname{GPT-4.1} & 58.17 & 55.28 & 55.47 & \colorbox{backblue!75}{73.33} & 37.50 & 21.43 & 27.27 & 78.85 & 89.13 & 83.67 \\
\bottomrule
\end{tabular}%
}
\end{table}

\begin{table}[t]
\centering
\caption{Performance comparison ($\%$) of perturbation direction analysis on scATAC-seq data. }
\label{tab:perturbation-direction}
\resizebox{\columnwidth}{!}{%
\begin{tabular}{l|cccc|ccc|ccc}
\toprule
\textbf{Setting} & \multicolumn{4}{c|}{\textbf{Overall}} & \multicolumn{3}{c|}{\textbf{Up}} & \multicolumn{3}{c}{\textbf{Down}} \\
\midrule
\textbf{Metric} & \textbf{Precision} & \textbf{Recall} & \textbf{F1} & \textbf{Accuracy} & \textbf{Precision} & \textbf{Recall} & \textbf{F1} & \textbf{Precision} & \textbf{Recall} & \textbf{F1} \\
\midrule
\rowcolor{black!10}\multicolumn{11}{c}{\textit{Open-source LLMs}} \\
\midrule
\llmname{Qwen-2.5-7B} & \colorbox{backred!50}{60.78} & 17.06 & 24.36 & 29.63 & \colorbox{backred!50}{82.35} & 41.18 & 54.90 & \colorbox{backred!50}{100.00} & 10.00 & 18.18 \\
\llmname{Qwen-2.5-32B} & 27.08 & 9.31 & 11.11 & 11.11 & 50.00 & 2.94 & 5.56 & 31.25 & 25.00 & 27.78 \\
\llmname{Qwen-2.5-72B} &38.10& 25.49& 29.17 &35.19 &64.29 &26.47 &37.50 &50.00 &50.00 & \colorbox{backblue!75}{50.00}  \\
\llmname{Llama-3.3-70B} &55.61 &\colorbox{backblue!75}{55.88}& \colorbox{backblue!75}{55.56} & 57.41 &67.74 & \colorbox{backblue!75}{61.76} &64.62 &43.48 &50.00 &46.51  \\
\llmname{DeepSeek-V3} & 35.95 & 16.86 & 22.94 & 24.07 & 41.18 & 20.59 & 27.45 & \colorbox{backblue!75}{66.67} & 30.00 & 41.38 \\
\llmname{DeepSeek-R1} & \colorbox{backblue!75}{58.75} & \colorbox{backred!50}{57.21} & \colorbox{backred!50}{57.07} & \colorbox{backred!50}{62.96} & 67.50 & \colorbox{backred!50}{79.41} & \colorbox{backred!50}{72.97} & 50.00 & 35.00 & 41.18 \\
\midrule
\rowcolor{black!10}\multicolumn{11}{c}{\textit{Closed-source LLMs}} \\
\midrule
\llmname{GPT-4} & 0.00 & 0.00 & 0.00 & 0.00 & 0.00 & 0.00 & 0.00 & 0.00 & 0.00 & 0.00 \\
\llmname{GPT-4o-mini} & 22.82 & 24.90 & 20.43 & 31.48 & 38.46 & 14.71 & 21.28 & 30.00 & \colorbox{backred!50}{60.00} & 40.00 \\
\llmname{GPT-4o} & 55.08 & 55.44 & 54.56 & 55.56 & 67.86 & 55.88 & 61.29 & 42.31 & \colorbox{backblue!75}{55.00} & 47.83 \\
\llmname{GPT-4.1-mini} & 43.18 & 40.59 & 41.63 & \colorbox{backblue!75}{61.11} & \colorbox{backblue!75}{75.00} & \colorbox{backblue!75}{61.76} & \colorbox{backblue!75}{67.74} & 54.55 & \colorbox{backred!50}{60.00} & \colorbox{backred!50}{57.14} \\
\llmname{GPT-4.1} & 49.13 & 49.12 & 46.28 & 46.30 & 61.90 & 38.24 & 47.27 & 36.36 & \colorbox{backred!50}{60.00} & 45.28\\
\bottomrule
\end{tabular}%
\vspace{-10mm}
}
\end{table}
\vspace{-1.5mm}

\begin{figure}[t]
\begin{subfigure}
    \centering
    \includegraphics[width=\linewidth]{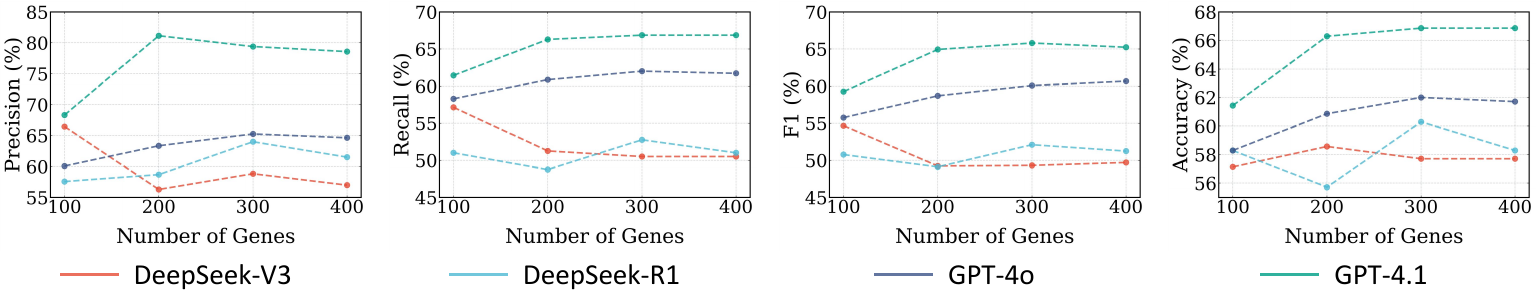}
    \vspace{-5mm}
    \caption{Performance comparison of cell type annotation when scaling context lengths.}
    \label{fig:context_length}
\end{subfigure}
\vspace{3mm}
\begin{subfigure}
    \centering
    \includegraphics[width=\linewidth]{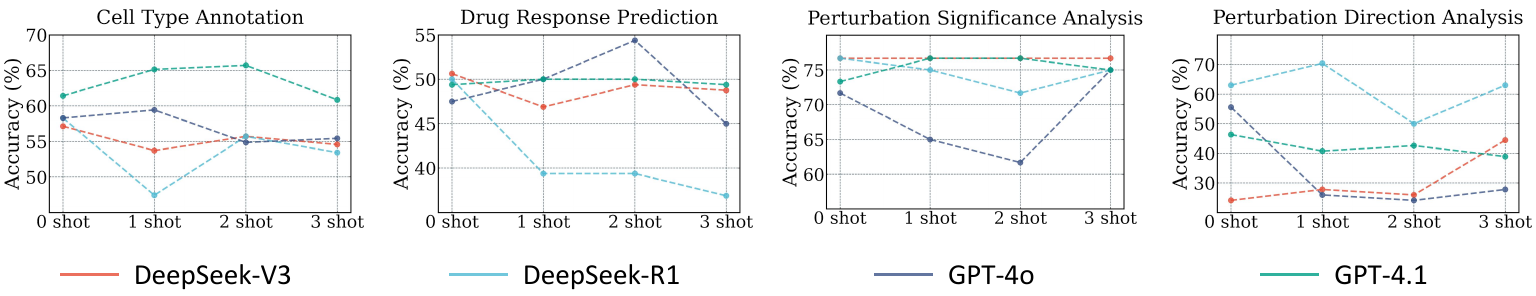}
    \vspace{-5mm}
    \caption{Performance comparison under few-shot settings across various tasks.}
    \label{fig:few_shot_acc}
\end{subfigure}
\vspace{3mm}
\begin{subfigure}
    \centering
    \includegraphics[width=\linewidth]{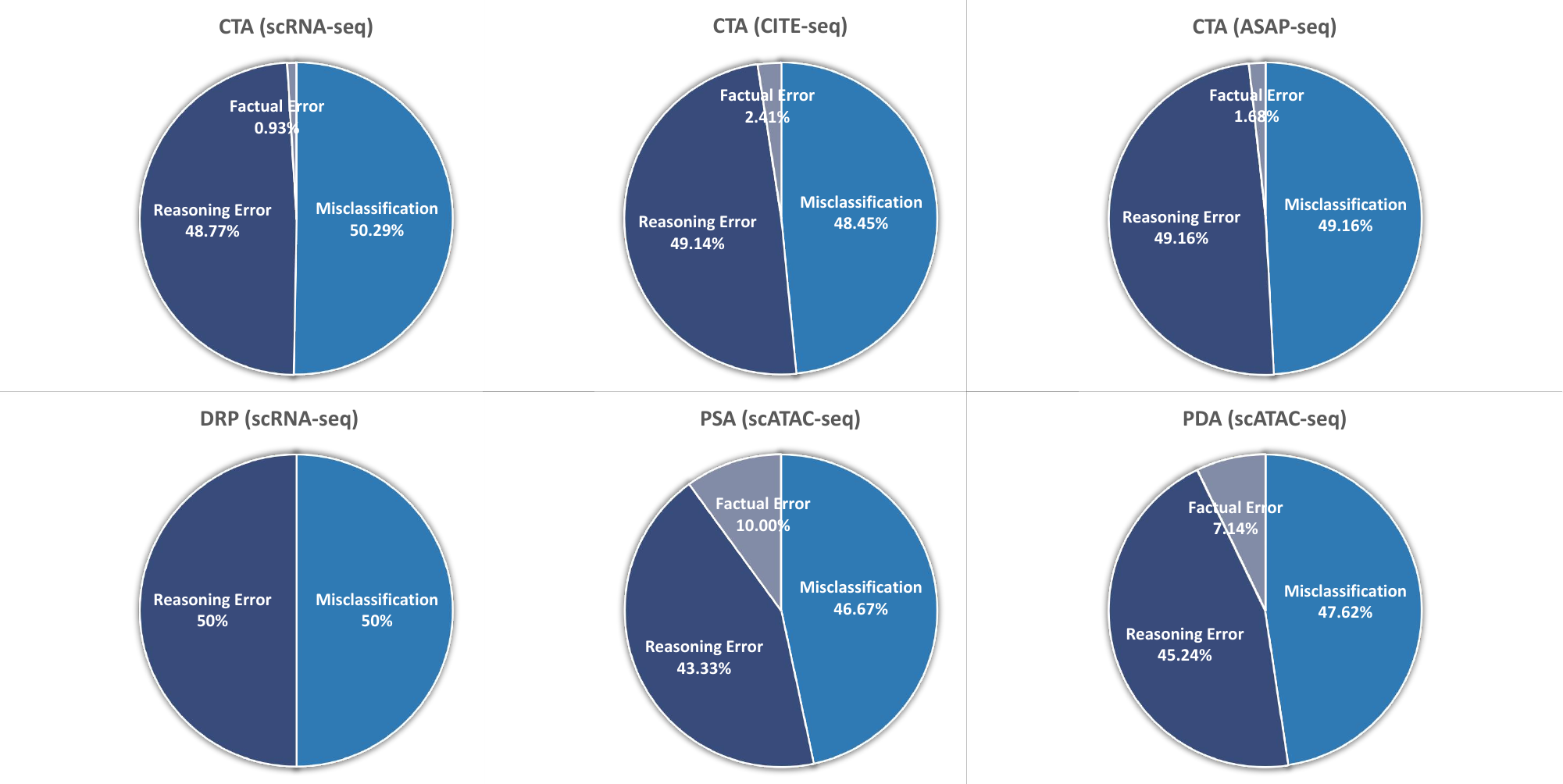}
    \vspace{-5mm}
    \caption{Distribution of \llmname{DeepSeek-R1}'s errors within distinct types across various tasks.}
    \label{fig:error_type_deepseek_r1}
\end{subfigure}
\vspace{-3mm}
\end{figure}


\subsection{Experimental Results and In-depth Analysis}
In Figure \ref{fig:leaderboard}, we present the leaderboard results for all sub-tasks. 
Since \llmname{C2S-Pythia-160M}, \llmname{C2S-Pythia-410M}, and \llmname{C2S-Pythia-1B} perform similarly across all tasks, we aggregate their results and refer to them collectively as \llmname{C2S-Pythia}. 
Detailed results for each sub-task—cell type annotation, drug response prediction, perturbation significance analysis, and perturbation direction analysis—are provided in Table \ref{tab:cta}, Table \ref{tab:drp}, Table 3 \ref{tab:perturbation-significance}, and Table \ref{tab:perturbation-direction}, respectively. 
The best and second-best results in the tables are marked in \colorbox{backred!50}{red} and \colorbox{backblue!75}{blue}, respectively.
More results and analysis can be found in Appendix \ref{sec:appendix_add_exp}, \ref{sec:appendix_error_ana}, and \ref{sec:appendix_case}.
From the results, we can derive the following observations.

\paragraph{Obs. 1: Generalist Models Perform Better than Specialist Models.}
Although the specialist models have been trained on language-centric single-cell analysis tasks, experimental results in Figure \ref{fig:leaderboard} show that \llmname{C2S-Pythia} fails to produce reliable predictions across all sub-tasks on \method{}. This suggests that, due to limitations in model capacity and training data, the specialist models may be overfitting rather than learning to generalize to unseen questions. In contrast, despite lacking task-specific training, larger-capacity generalist models exhibit emerging capabilities in reasoning about cell biology. These findings highlight the potential of leveraging strong generalist models as base architectures for future research in language-driven single-cell analysis.

\paragraph{Obs. 2: LLM Performance Scales with Model Size.}
We observe that \llmname{GPT-4} and \llmname{DeepSeek} family models generally outperform the \llmname{Llama} and \llmname{Qwen} series across all tasks. This trend suggests a positive correlation between model capacity and performance in cell biology understanding. Notably, across the six task-specific leaderboards, all top-performing models belong to either the \llmname{DeepSeek} or \llmname{GPT-4} families—securing four and two first-place rankings, respectively. These results indicate that the scaling laws of LLMs also hold in the context of cell biology.

\paragraph{Obs. 3: Current LLMs Demonstrate Limited Understanding of Cell Biology.}
While LLMs exhibit preliminary capabilities for cell biology, their overall performance remains far from satisfactory. On the cell type annotation task, the top-performing models achieved accuracies of only $42.38\%$, $61.43\%$, and $29.33\%$ on scRNA-seq, CITE-seq, and ASAP-seq data, respectively. For the binary drug response prediction task, the best model achieved only $55\%$ accuracy, indicating no substantial improvement over random guessing. In the perturbation analysis tasks, we observe that LLMs perform better in assessing perturbation significance than in predicting perturbation direction, which aligns with intuitive expectations. Most models also fail to outperform random guessing in the direction prediction task. \llmname{GPT-4} often refuses to answer for both drug response prediction and perturbation analysis tasks, while \llmname{Qwen} family models tend to answer "No" to all questions in the perturbation significance analysis task. Overall, these results highlight that current LLMs are far from satisfactory and there remains substantial room for improvement in LLMs for cell biology.

\paragraph{Obs. 4: Scaling Context Lengths May Not Consistently Improve Performance.}
Figure \ref{fig:context_length} presents the cell type annotation results of four advanced LLMs on CITE-seq data as the context length increases.
We scale the context length by including more gene names in the cell sentences.
For the \llmname{GPT-4} family, performance generally improves with longer context lengths, suggesting these models benefit from richer input information. 
In contrast, for the \llmname{DeepSeek} family, increasing context length does not consistently lead to performance gains. We hypothesize that this is because \llmname{DeepSeek} models already demonstrate strong reasoning capabilities with relatively short contexts, and adding more genes with low expressions may introduce noise that hinders model performance.

\paragraph{Obs. 5: Few-shot In-context Learning Does Not Always Boost Performance.}
Unlike general tasks, few-shot in-context learning yields limited gains and often even degrades performance in single-cell analysis. Results in Figure \ref{fig:few_shot_acc} exhibit no consistent improvement. We hypothesize that this is largely due to the high level of noise inherent in single-cell data, which hampers the model’s ability to generalize when guided by noisy examples. This observation highlights an important insight: In few-shot settings, sample quality may be more critical than sample quantity.

\paragraph{Obs. 6: Error Profiles Vary across Tasks.}
Figure~\ref{fig:error_type_deepseek_r1} shows the distribution of error types made by \llmname{DeepSeek-R1} across different tasks. We observe that the most frequent error categories are reasoning errors and misclassifications. Additionally, another type of error, factual errors, is more prominent in gene-level tasks, whereas they are relatively rare in cell-level and drug-level tasks.

\section{Conclusion}\label{sec:conclusion}

In this work, we introduced \method{}, a unified language-centric benchmark that encompasses single-cell multi-omics data and spans hierarchical single-cell analysis tasks, addressing the absence of systematic evaluation in LLMs for cell biology. Through a comprehensive assessment of 14 advanced LLMs, we uncovered both their capabilities and limitations in reasoning over single-cell analysis tasks, laying the groundwork for future research within this domain.
In future works, we plan to (1) enhance the scalability and diversity of \method{}, and (2) build on \method{} to advance natural language understanding as a next-generation paradigm for interpreting cell biology.

\bibliography{rec.bib}

\begin{thebibliography}{10}

\bibitem{achiam2023gpt}
Josh Achiam, Steven Adler, Sandhini Agarwal, Lama Ahmad, Ilge Akkaya, Florencia~Leoni Aleman, Diogo Almeida, Janko Altenschmidt, Sam Altman, Shyamal Anadkat, et~al.
\newblock Gpt-4 technical report.
\newblock {\em arXiv preprint arXiv:2303.08774}, 2023.

\bibitem{adam2020machine}
George Adam, Ladislav Ramp{\'a}{\v{s}}ek, Zhaleh Safikhani, Petr Smirnov, Benjamin Haibe-Kains, and Anna Goldenberg.
\newblock Machine learning approaches to drug response prediction: challenges and recent progress.
\newblock {\em NPJ precision oncology}, 4(1):19, 2020.

\bibitem{adamson2016multiplexed}
Britt Adamson, Thomas~M Norman, Marco Jost, Min~Y Cho, James~K Nu{\~n}ez, Yuwen Chen, Jacqueline~E Villalta, Luke~A Gilbert, Max~A Horlbeck, Marco~Y Hein, et~al.
\newblock A multiplexed single-cell crispr screening platform enables systematic dissection of the unfolded protein response.
\newblock {\em Cell}, 167(7):1867--1882, 2016.

\bibitem{ahn2024large}
Janice Ahn, Rishu Verma, Renze Lou, Di~Liu, Rui Zhang, and Wenpeng Yin.
\newblock Large language models for mathematical reasoning: Progresses and challenges.
\newblock {\em arXiv preprint arXiv:2402.00157}, 2024.

\bibitem{aissa2021single}
Alexandre~F Aissa, Abul~BMMK Islam, Majd~M Ariss, Cammille~C Go, Alexandra~E Rader, Ryan~D Conrardy, Alexa~M Gajda, Carlota Rubio-Perez, Klara Valyi-Nagy, Mary Pasquinelli, et~al.
\newblock Single-cell transcriptional changes associated with drug tolerance and response to combination therapies in cancer.
\newblock {\em Nature communications}, 12(1):1628, 2021.

\bibitem{argelaguet2020mofa+}
Ricard Argelaguet, Damien Arnol, Danila Bredikhin, Yonatan Deloro, Britta Velten, John~C Marioni, and Oliver Stegle.
\newblock Mofa+: a statistical framework for comprehensive integration of multi-modal single-cell data.
\newblock {\em Genome biology}, 21(1):1--17, 2020.

\bibitem{baruzzo2020sparsim}
Giacomo Baruzzo, Ilaria Patuzzi, and Barbara Di~Camillo.
\newblock Sparsim single cell: a count data simulator for scrna-seq data.
\newblock {\em Bioinformatics}, 36(5):1468--1475, 2020.

\bibitem{beltagy2019scibert}
Iz~Beltagy, Kyle Lo, and Arman Cohan.
\newblock Scibert: A pretrained language model for scientific text.
\newblock {\em arXiv preprint arXiv:1903.10676}, 2019.

\bibitem{brandow2022advances}
AM~Brandow and RI~Liem.
\newblock Advances in the diagnosis and treatment of sickle cell disease.
\newblock {\em Journal of Hematology \& Oncology}, 15(1):20, 2022.

\bibitem{brennecke2013accounting}
Philip Brennecke, Simon Anders, Jong~Kyoung Kim, Aleksandra~A Ko{\l}odziejczyk, Xiuwei Zhang, Valentina Proserpio, Bianka Baying, Vladimir Benes, Sarah~A Teichmann, John~C Marioni, et~al.
\newblock Accounting for technical noise in single-cell rna-seq experiments.
\newblock {\em Nature methods}, 10(11):1093--1095, 2013.

\bibitem{buettner2015computational}
Florian Buettner, Kedar~N Natarajan, F~Paolo Casale, Valentina Proserpio, Antonio Scialdone, Fabian~J Theis, Sarah~A Teichmann, John~C Marioni, and Oliver Stegle.
\newblock Computational analysis of cell-to-cell heterogeneity in single-cell rna-sequencing data reveals hidden subpopulations of cells.
\newblock {\em Nature biotechnology}, 33(2):155--160, 2015.

\bibitem{cai2024sciassess}
Hengxing Cai, Xiaochen Cai, Junhan Chang, Sihang Li, Lin Yao, Changxin Wang, Zhifeng Gao, Hongshuai Wang, Yongge Li, Mujie Lin, et~al.
\newblock Sciassess: Benchmarking llm proficiency in scientific literature analysis.
\newblock {\em arXiv preprint arXiv:2403.01976}, 2024.

\bibitem{chen2021geoqa}
Jiaqi Chen, Jianheng Tang, Jinghui Qin, Xiaodan Liang, Lingbo Liu, Eric~P Xing, and Liang Lin.
\newblock Geoqa: A geometric question answering benchmark towards multimodal numerical reasoning.
\newblock {\em arXiv preprint arXiv:2105.14517}, 2021.

\bibitem{cui2024scgpt}
Haotian Cui, Chloe Wang, Hassaan Maan, Kuan Pang, Fengning Luo, Nan Duan, and Bo~Wang.
\newblock scgpt: toward building a foundation model for single-cell multi-omics using generative ai.
\newblock {\em Nature Methods}, 21(8):1470--1480, 2024.

\bibitem{cuzick1985wilcoxon}
Jack Cuzick.
\newblock A wilcoxon-type test for trend.
\newblock {\em Statistics in medicine}, 4(1):87--90, 1985.

\bibitem{davidson2005gene}
Eric Davidson and Michael Levin.
\newblock Gene regulatory networks.
\newblock {\em Proceedings of the National Academy of Sciences}, 102(14):4935--4935, 2005.

\bibitem{gehan1965generalized}
Edmund~A Gehan.
\newblock A generalized wilcoxon test for comparing arbitrarily singly-censored samples.
\newblock {\em Biometrika}, 52(1-2):203--224, 1965.

\bibitem{grattafiori2024llama}
Aaron Grattafiori, Abhimanyu Dubey, Abhinav Jauhri, Abhinav Pandey, Abhishek Kadian, Ahmad Al-Dahle, Aiesha Letman, Akhil Mathur, Alan Schelten, Alex Vaughan, et~al.
\newblock The llama 3 herd of models.
\newblock {\em arXiv preprint arXiv:2407.21783}, 2024.

\bibitem{guo2025deepseek}
Daya Guo, Dejian Yang, Haowei Zhang, Junxiao Song, Ruoyu Zhang, Runxin Xu, Qihao Zhu, Shirong Ma, Peiyi Wang, Xiao Bi, et~al.
\newblock Deepseek-r1: Incentivizing reasoning capability in llms via reinforcement learning.
\newblock {\em arXiv preprint arXiv:2501.12948}, 2025.

\bibitem{hao2024large}
Minsheng Hao, Jing Gong, Xin Zeng, Chiming Liu, Yucheng Guo, Xingyi Cheng, Taifeng Wang, Jianzhu Ma, Xuegong Zhang, and Le~Song.
\newblock Large-scale foundation model on single-cell transcriptomics.
\newblock {\em Nature methods}, 21(8):1481--1491, 2024.

\bibitem{heath2016single}
James~R Heath, Antoni Ribas, and Paul~S Mischel.
\newblock Single-cell analysis tools for drug discovery and development.
\newblock {\em Nature reviews Drug discovery}, 15(3):204--216, 2016.

\bibitem{hendrycks2021measuring}
Dan Hendrycks, Collin Burns, Saurav Kadavath, Akul Arora, Steven Basart, Eric Tang, Dawn Song, and Jacob Steinhardt.
\newblock Measuring mathematical problem solving with the math dataset.
\newblock {\em arXiv preprint arXiv:2103.03874}, 2021.

\bibitem{heumos2023best}
Lukas Heumos, Anna~C Schaar, Christopher Lance, Anastasia Litinetskaya, Felix Drost, Luke Zappia, Malte~D L{\"u}cken, Daniel~C Strobl, Juan Henao, Fabiola Curion, et~al.
\newblock Best practices for single-cell analysis across modalities.
\newblock {\em Nature Reviews Genetics}, 24(8):550--572, 2023.

\bibitem{imani2023mathprompter}
Shima Imani, Liang Du, and Harsh Shrivastava.
\newblock Mathprompter: Mathematical reasoning using large language models.
\newblock {\em arXiv preprint arXiv:2303.05398}, 2023.

\bibitem{ji2021machine}
Yuge Ji, Mohammad Lotfollahi, F~Alexander Wolf, and Fabian~J Theis.
\newblock Machine learning for perturbational single-cell omics.
\newblock {\em Cell Systems}, 12(6):522--537, 2021.

\bibitem{jin2019pubmedqa}
Qiao Jin, Bhuwan Dhingra, Zhengping Liu, William~W Cohen, and Xinghua Lu.
\newblock Pubmedqa: A dataset for biomedical research question answering.
\newblock {\em arXiv preprint arXiv:1909.06146}, 2019.

\bibitem{june2018car}
Carl~H June, Roddy~S O’Connor, Omkar~U Kawalekar, Saba Ghassemi, and Michael~C Milone.
\newblock Car t cell immunotherapy for human cancer.
\newblock {\em Science}, 359(6382):1361--1365, 2018.

\bibitem{karlebach2008modelling}
Guy Karlebach and Ron Shamir.
\newblock Modelling and analysis of gene regulatory networks.
\newblock {\em Nature reviews Molecular cell biology}, 9(10):770--780, 2008.

\bibitem{korthauer2016statistical}
Keegan~D Korthauer, Li-Fang Chu, Michael~A Newton, Yuan Li, James Thomson, Ron Stewart, and Christina Kendziorski.
\newblock A statistical approach for identifying differential distributions in single-cell rna-seq experiments.
\newblock {\em Genome biology}, 17:1--15, 2016.

\bibitem{kwon2023efficient}
Woosuk Kwon, Zhuohan Li, Siyuan Zhuang, Ying Sheng, Lianmin Zheng, Cody~Hao Yu, Joseph~E. Gonzalez, Hao Zhang, and Ion Stoica.
\newblock Efficient memory management for large language model serving with pagedattention.
\newblock In {\em Proceedings of the ACM SIGOPS 29th Symposium on Operating Systems Principles}, 2023.

\bibitem{lecun2015deep}
Yann LeCun, Yoshua Bengio, and Geoffrey Hinton.
\newblock Deep learning.
\newblock {\em nature}, 521(7553):436--444, 2015.

\bibitem{lee2020single}
Jeongwoo Lee, Do~Young Hyeon, and Daehee Hwang.
\newblock Single-cell multiomics: technologies and data analysis methods.
\newblock {\em Experimental \& Molecular Medicine}, 52(9):1428--1442, 2020.

\bibitem{lee2020biobert}
Jinhyuk Lee, Wonjin Yoon, Sungdong Kim, Donghyeon Kim, Sunkyu Kim, Chan~Ho So, and Jaewoo Kang.
\newblock Biobert: a pre-trained biomedical language representation model for biomedical text mining.
\newblock {\em Bioinformatics}, 36(4):1234--1240, 2020.

\bibitem{levine2024cell2sentence}
Daniel Levine, Syed~Asad Rizvi, Sacha L{\'e}vy, Nazreen Pallikkavaliyaveetil, David Zhang, Xingyu Chen, Sina Ghadermarzi, Ruiming Wu, Zihe Zheng, Ivan Vrkic, et~al.
\newblock Cell2sentence: teaching large language models the language of biology.
\newblock {\em BioRxiv}, pages 2023--09, 2024.

\bibitem{li2020deep}
Xiangjie Li, Kui Wang, Yafei Lyu, Huize Pan, Jingxiao Zhang, Dwight Stambolian, Katalin Susztak, Muredach~P Reilly, Gang Hu, and Mingyao Li.
\newblock Deep learning enables accurate clustering with batch effect removal in single-cell rna-seq analysis.
\newblock {\em Nature communications}, 11(1):2338, 2020.

\bibitem{li2022competition}
Yujia Li, David Choi, Junyoung Chung, Nate Kushman, Julian Schrittwieser, R{\'e}mi Leblond, Tom Eccles, James Keeling, Felix Gimeno, Agustin Dal~Lago, et~al.
\newblock Competition-level code generation with alphacode.
\newblock {\em Science}, 378(6624):1092--1097, 2022.

\bibitem{lindstrom2010overview}
Sara Lindstr{\"o}m and Helene Andersson-Svahn.
\newblock Overview of single-cell analyses: microdevices and applications.
\newblock {\em Lab on a Chip}, 10(24):3363--3372, 2010.

\bibitem{liu2024deepseek}
Aixin Liu, Bei Feng, Bing Xue, Bingxuan Wang, Bochao Wu, Chengda Lu, Chenggang Zhao, Chengqi Deng, Chenyu Zhang, Chong Ruan, et~al.
\newblock Deepseek-v3 technical report.
\newblock {\em arXiv preprint arXiv:2412.19437}, 2024.

\bibitem{lu2023mathvista}
Pan Lu, Hritik Bansal, Tony Xia, Jiacheng Liu, Chunyuan Li, Hannaneh Hajishirzi, Hao Cheng, Kai-Wei Chang, Michel Galley, and Jianfeng Gao.
\newblock Mathvista: Evaluating mathematical reasoning of foundation models in visual contexts.
\newblock {\em arXiv preprint arXiv:2310.02255}, 2023.

\bibitem{luo2022biogpt}
Renqian Luo, Liai Sun, Yingce Xia, Tao Qin, Sheng Zhang, Hoifung Poon, and Tie-Yan Liu.
\newblock Biogpt: generative pre-trained transformer for biomedical text generation and mining.
\newblock {\em Briefings in bioinformatics}, 23(6):bbac409, 2022.

\bibitem{ma2022deep}
Qin Ma and Dong Xu.
\newblock Deep learning shapes single-cell data analysis.
\newblock {\em Nature reviews Molecular cell biology}, 23(5):303--304, 2022.

\bibitem{madani2023large}
Ali Madani, Ben Krause, Eric~R Greene, Subu Subramanian, Benjamin~P Mohr, James~M Holton, Jose~Luis Olmos~Jr, Caiming Xiong, Zachary~Z Sun, Richard Socher, et~al.
\newblock Large language models generate functional protein sequences across diverse families.
\newblock {\em Nature biotechnology}, 41(8):1099--1106, 2023.

\bibitem{mimitou2021scalable}
Eleni~P Mimitou, Caleb~A Lareau, Kelvin~Y Chen, Andre~L Zorzetto-Fernandes, Yuhan Hao, Yusuke Takeshima, Wendy Luo, Tse-Shun Huang, Bertrand~Z Yeung, Efthymia Papalexi, et~al.
\newblock Scalable, multimodal profiling of chromatin accessibility, gene expression and protein levels in single cells.
\newblock {\em Nature biotechnology}, 39(10):1246--1258, 2021.

\bibitem{openaigpt4omini}
OpenAI.
\newblock Gpt-4o-mini system card, 2024.

\bibitem{openaigpt4o}
OpenAI.
\newblock Gpt-4o system card, 2024.

\bibitem{openaigpt41}
OpenAI.
\newblock Gpt-4.1 system card, 2025.

\bibitem{pan2023large}
Jie Pan.
\newblock Large language model for molecular chemistry.
\newblock {\em Nature Computational Science}, 3(1):5--5, 2023.

\bibitem{pasquini2021automated}
Giovanni Pasquini, Jesus Eduardo~Rojo Arias, Patrick Sch{\"a}fer, and Volker Busskamp.
\newblock Automated methods for cell type annotation on scrna-seq data.
\newblock {\em Computational and Structural Biotechnology Journal}, 19:961--969, 2021.

\bibitem{picelli2014full}
Simone Picelli, Omid~R Faridani, {\AA}sa~K Bj{\"o}rklund, G{\"o}sta Winberg, Sven Sagasser, and Rickard Sandberg.
\newblock Full-length rna-seq from single cells using smart-seq2.
\newblock {\em Nature protocols}, 9(1):171--181, 2014.

\bibitem{rizvi2025scaling}
Syed~Asad Rizvi, Daniel Levine, Aakash Patel, Shiyang Zhang, Eric Wang, Sizhuang He, David Zhang, Cerise Tang, Zhuoyang Lyu, Rayyan Darji, et~al.
\newblock Scaling large language models for next-generation single-cell analysis.
\newblock {\em bioRxiv}, pages 2025--04, 2025.

\bibitem{schirmer2019neuronal}
Lucas Schirmer, Dmitry Velmeshev, Staffan Holmqvist, Max Kaufmann, Sebastian Werneburg, Diane Jung, Stephanie Vistnes, John~H Stockley, Adam Young, Maike Steindel, et~al.
\newblock Neuronal vulnerability and multilineage diversity in multiple sclerosis.
\newblock {\em Nature}, 573(7772):75--82, 2019.

\bibitem{sun2024scieval}
Liangtai Sun, Yang Han, Zihan Zhao, Da~Ma, Zhennan Shen, Baocai Chen, Lu~Chen, and Kai Yu.
\newblock Scieval: A multi-level large language model evaluation benchmark for scientific research.
\newblock In {\em Proceedings of the AAAI Conference on Artificial Intelligence}, volume~38, pages 19053--19061, 2024.

\bibitem{taylor2022galactica}
Ross Taylor, Marcin Kardas, Guillem Cucurull, Thomas Scialom, Anthony Hartshorn, Elvis Saravia, Andrew Poulton, Viktor Kerkez, and Robert Stojnic.
\newblock Galactica: A large language model for science.
\newblock {\em arXiv preprint arXiv:2211.09085}, 2022.

\bibitem{vaswani2017attention}
Ashish Vaswani, Noam Shazeer, Niki Parmar, Jakob Uszkoreit, Llion Jones, Aidan~N Gomez, {\L}ukasz Kaiser, and Illia Polosukhin.
\newblock Attention is all you need.
\newblock {\em Advances in neural information processing systems}, 30, 2017.

\bibitem{wang2023scibench}
Xiaoxuan Wang, Ziniu Hu, Pan Lu, Yanqiao Zhu, Jieyu Zhang, Satyen Subramaniam, Arjun~R Loomba, Shichang Zhang, Yizhou Sun, and Wei Wang.
\newblock Scibench: Evaluating college-level scientific problem-solving abilities of large language models.
\newblock {\em arXiv preprint arXiv:2307.10635}, 2023.

\bibitem{wei2022chain}
Jason Wei, Xuezhi Wang, Dale Schuurmans, Maarten Bosma, Fei Xia, Ed~Chi, Quoc~V Le, Denny Zhou, et~al.
\newblock Chain-of-thought prompting elicits reasoning in large language models.
\newblock {\em Advances in neural information processing systems}, 35:24824--24837, 2022.

\bibitem{xu2023symbol}
Fangzhi Xu, Zhiyong Wu, Qiushi Sun, Siyu Ren, Fei Yuan, Shuai Yuan, Qika Lin, Yu~Qiao, and Jun Liu.
\newblock Symbol-llm: Towards foundational symbol-centric interface for large language models.
\newblock {\em arXiv preprint arXiv:2311.09278}, 2023.

\bibitem{yang2024qwen2}
An~Yang, Baosong Yang, Beichen Zhang, Binyuan Hui, Bo~Zheng, Bowen Yu, Chengyuan Li, Dayiheng Liu, Fei Huang, Haoran Wei, et~al.
\newblock Qwen2. 5 technical report.
\newblock {\em arXiv preprint arXiv:2412.15115}, 2024.

\bibitem{zheng2025large}
Yizhen Zheng, Huan~Yee Koh, Jiaxin Ju, Anh~TN Nguyen, Lauren~T May, Geoffrey~I Webb, and Shirui Pan.
\newblock Large language models for scientific discovery in molecular property prediction.
\newblock {\em Nature Machine Intelligence}, pages 1--11, 2025.

\end{thebibliography}
\bibliographystyle{plain}

\clearpage
\appendix

\section*{Appendix Overview}
\begin{itemize}[leftmargin=*]
    \item Section~\ref{sec:appendix_cellverse_details}: More Details about \method{}.
    \item Section~\ref{sec:appendix_add_exp}: Additional Experimental Results.
    \item Section~\ref{sec:appendix_error_ana}: Additional Error Analysis.
    \item Section~\ref{sec:appendix_case}: Case Study.
    \item Section~\ref{sec:appendix_limit}: Limitation and Future Work.
\end{itemize}

\section{More Details about \method{}}\label{sec:appendix_cellverse_details}

\subsection{Model Sources}
For all evaluated LLMs, we select their latest versions and best-performing configurations to accurately assess their capabilities in understanding cell biology. Table~\ref{tab:model_scource} summarizes the release dates and sources of the models included in \method{}.

\begin{table}[h]
    \centering
    \caption{The release dates and sources of the evaluated models in \method{}.}
    \label{tab:model_scource}
    \resizebox{\linewidth}{!}{ 
    \begin{tabular}{l@{\hspace{0.5cm}}c@{\hspace{1cm}}p{0.7\textwidth}}
    \toprule
    \textbf{Model} & \textbf{\makecell{Release Date}} & \textbf{\makecell[c]{Source}} \\
    \midrule
    \llmname{C2S-Pythia-160M} \cite{levine2024cell2sentence}   &   2024-02    & \url{https://huggingface.co/vandijklab/pythia-160m-c2s} \\
    \midrule
    \multirow{2}{*}{\llmname{C2S-Pythia-410M}} \cite{levine2024cell2sentence}  &   \multirow{2}{*}{2024-09}    & \url{https://huggingface.co/vandijklab/C2S-Pythia-410m-diverse-single-and-multi-cell-tasks} \\
    \midrule
    \multirow{2}{*}{\llmname{C2S-Pythia-1B}} \cite{rizvi2025scaling}   &   \multirow{2}{*}{2025-04}    & \url{https://huggingface.co/vandijklab/C2S-Scale-Pythia-1b-pt} \\
    \midrule
    \llmname{Qwen-2.5-7B} \cite{yang2024qwen2}  &   2024-09   & \url{https://huggingface.co/Qwen/Qwen2.5-7B} \\
    \midrule
    \llmname{Qwen-2.5-32B} \cite{yang2024qwen2} &   2024-09   & \url{https://huggingface.co/Qwen/Qwen2.5-32B} \\
    \midrule
    \llmname{Qwen-2.5-72B} \cite{yang2024qwen2}  &   2024-09    & \url{https://huggingface.co/Qwen/Qwen2.5-72B}\\
    \midrule
    \multirow{2}{*}{\llmname{Llama-3.3-70B} \cite{grattafiori2024llama}}   &   \multirow{2}{*}{2024-12}    & \url{https://huggingface.co/meta-llama/Llama-3.3-70B-Instruct}\\
    \midrule
    \llmname{DeepSeek-V3 \cite{liu2024deepseek}}   &   2025-03    & \url{https://huggingface.co/deepseek-ai/DeepSeek-V3}\\
     \midrule
    \llmname{DeepSeek-R1} \cite{guo2025deepseek} &  2025-03    & \url{https://huggingface.co/deepseek-ai/DeepSeek-R1}\\
    \midrule
    \llmname{GPT-4} \cite{achiam2023gpt}  &   2023-06    & \url{https://platform.openai.com/docs/models/gpt-4}\\
    \midrule
    \llmname{GPT-4o-mini} \cite{openaigpt4omini}  &   2024-07    & \url{https://platform.openai.com/docs/models/gpt-4o-mini}\\
    \midrule
    \llmname{GPT-4o} \cite{openaigpt4o}   &   2024-11    & \url{https://platform.openai.com/docs/models/gpt-4o}\\
    \midrule
    \llmname{GPT-4.1-mini} \cite{openaigpt41}  &   2025-04    & \url{https://platform.openai.com/docs/models/gpt-4.1-mini}\\
    \midrule
    \llmname{GPT-4.1} \cite{openaigpt41}  &   2025-04    & \url{https://platform.openai.com/docs/models/gpt-4.1}\\
    \bottomrule
    \end{tabular}
    }
\end{table}

\subsection{Evaluation Prompts for Single-cell Analysis Tasks}
We design task-specific prompts to evaluate the performance of LLMs on different single-cell analysis tasks within \method{}. At the beginning of each interaction, we add a unified system prompt instructing the LLMs to act as experts in cell biology and genomics. We then integrate the transformed language-centric information with the task-specific prompts to form the final questions. Since current LLMs struggle to produce accurate predictions in open-ended formats, we convert all questions into a closed-set, multiple-choice setting to ensure more reliable evaluation. As shown in Table \ref{tab:prompt}, we summarize the prompt design strategies used for each task.

\begin{table}[h]
    \centering
    \caption{Evaluation prompt of LLMs for different single-cell analysis tasks.}
    \label{tab:prompt}
    \begin{tabular}{p{0.36\textwidth}p{0.52\textwidth}}
    \toprule
    \textbf{Task}                                       & \textbf{Prompt} \\
    \midrule
    \multirow{11}{*}{Cell Type Annotation}  & You are an expert who knows a lot about single cell biology and genomics and will help me solve a series of tasks related to single cell data analysis. Below is a list of \{gene\_names\} ordered by descending expression level in a single-cell sample. Your task is to select one cell type to which this cell most likely belongs from the following candidate cell types: \{cell\_types\}. Make your choice in format `Final Answer: Prediction'.\\
    &-- \textbf{Gene Names:} \{gene\_names\}\\
    &-- \textbf{Cell Types:} \{cell\_types\}\\
    \midrule
    \multirow{10}{*}{Drug Response Prediction}  & You are an expert who knows a lot about single cell biology and genomics and will help me solve a series of tasks related to single cell data analysis. Below is a list of \{gene\_names\} ordered by descending expression level in a cell. Your task is to predict its response to the drug \{drug\}: Responses: \{responses\}. Make your choice in format `Final Answer: Response'.\\
    &-- \textbf{Gene Names:} \{gene\_names\}\\
    &-- \textbf{Drug:} \{drug\}\\
    &-- \textbf{Responses:} \{responses\}\\
    \midrule
    \multirow{9}{*}{Perturbation Significance Analysis}  & You are an expert who knows a lot about single cell biology and genomics and will help me solve a series of tasks related to single cell data analysis. Does perturbing \{gene\_a\} in K562 cells cause significant changes in the expression of \{gene\_b\}? Make your choice in format `Final Answer: Yes' or `Final Answer: No'.\\
    &-- \textbf{Gene A:} \{gene\_a\}\\
    &-- \textbf{Gene B:} \{gene\_b\}\\
    \midrule
    \multirow{9}{*}{Perturbation Direction Analysis}  & You are an expert who knows a lot about single cell biology and genomics and will help me solve a series of tasks related to single cell data analysis. 
    Does perturbing \{gene\_a\} in K562 cells cause the expression of \{gene\_b\} to upregulate or downregulate? Make your answer in format `Final Answer: Up' or `Final Answer: Down'.\\
    &-- \textbf{Gene A:} \{gene\_a\}\\
    &-- \textbf{Gene B:} \{gene\_b\}\\
    \bottomrule
    \end{tabular}
\end{table}

\section{Additional Experimental Results}\label{sec:appendix_add_exp}

\subsection{Performance Comparison of Specific Cell Types}
\begin{figure}[h]
    \centering
    \includegraphics[width=\linewidth]{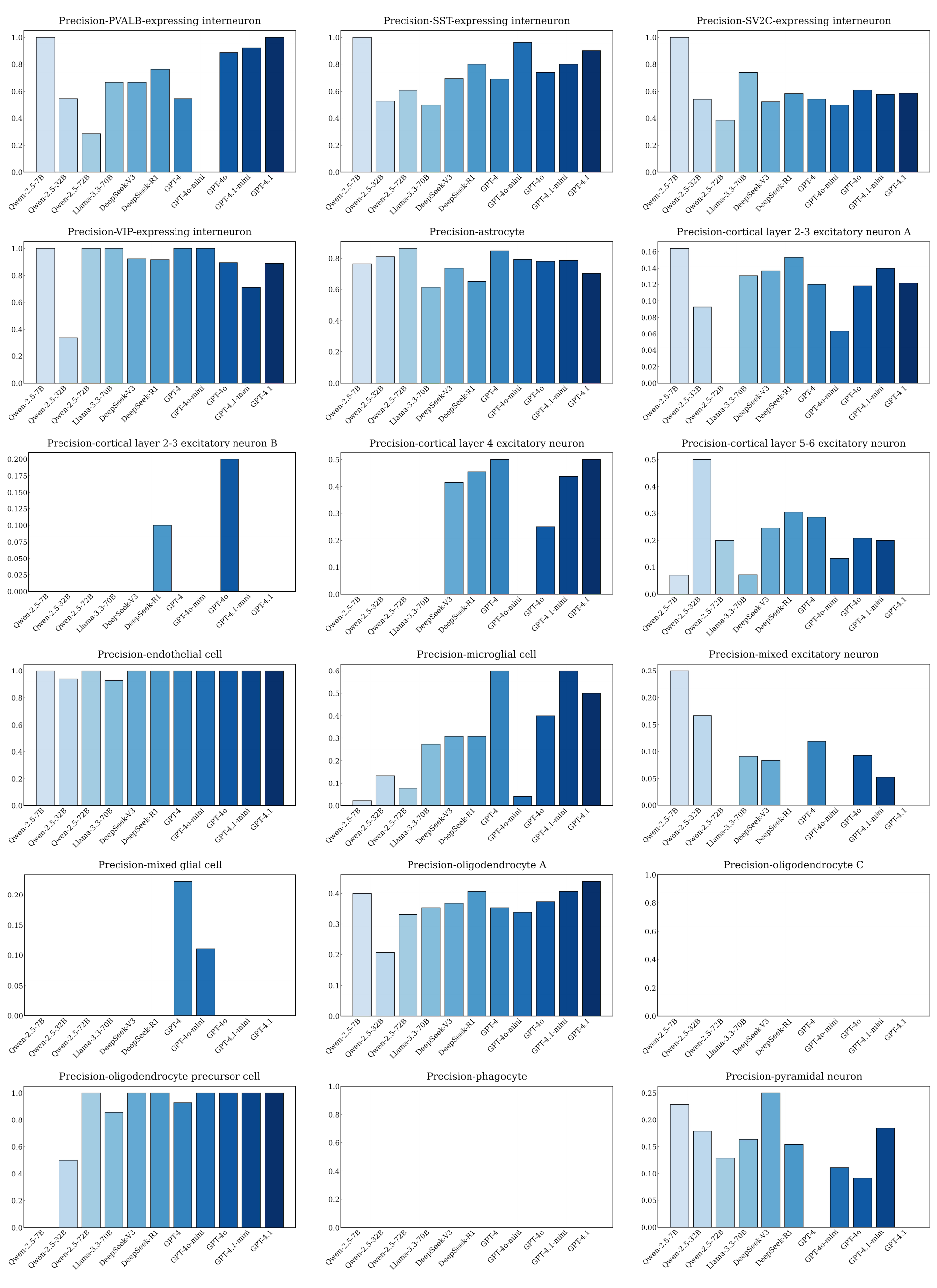}
    \caption{Precision score comparison of specific cell types on scRNA-seq data. All LLMs fail to predict certain challenging cell types, such as oligodendrocyte C and phagocyte.}
    \label{fig:precision_ms}
\end{figure}

\begin{figure}[h]
    \centering
    \includegraphics[width=\linewidth]{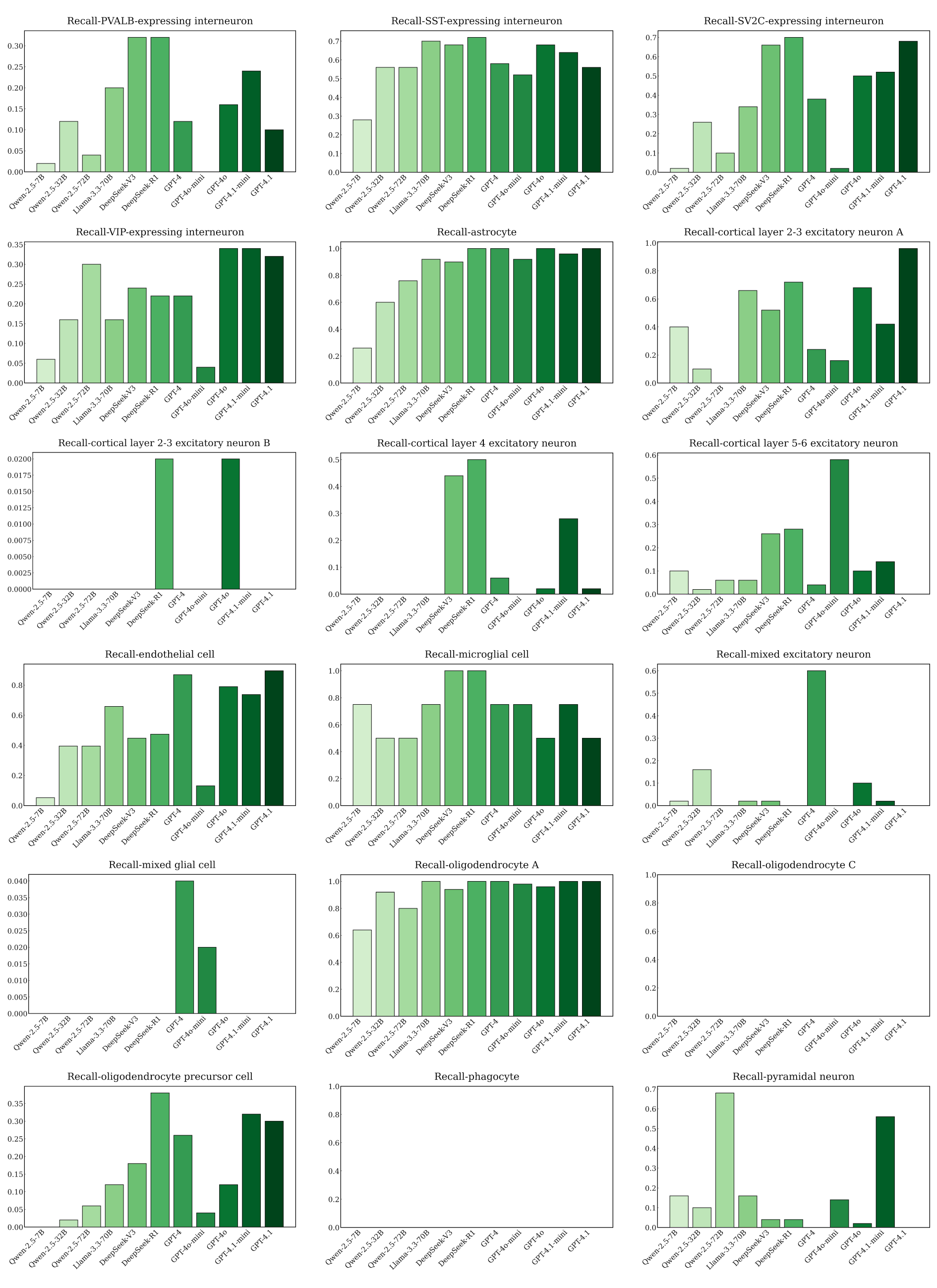}
    \caption{Recall score comparison of specific cell types on scRNA-seq data. All LLMs fail to predict certain challenging cell types, such as oligodendrocyte C and phagocyte.}
    \label{fig:recall_ms}
\end{figure}

\begin{figure}[h]
    \centering
    \includegraphics[width=\linewidth]{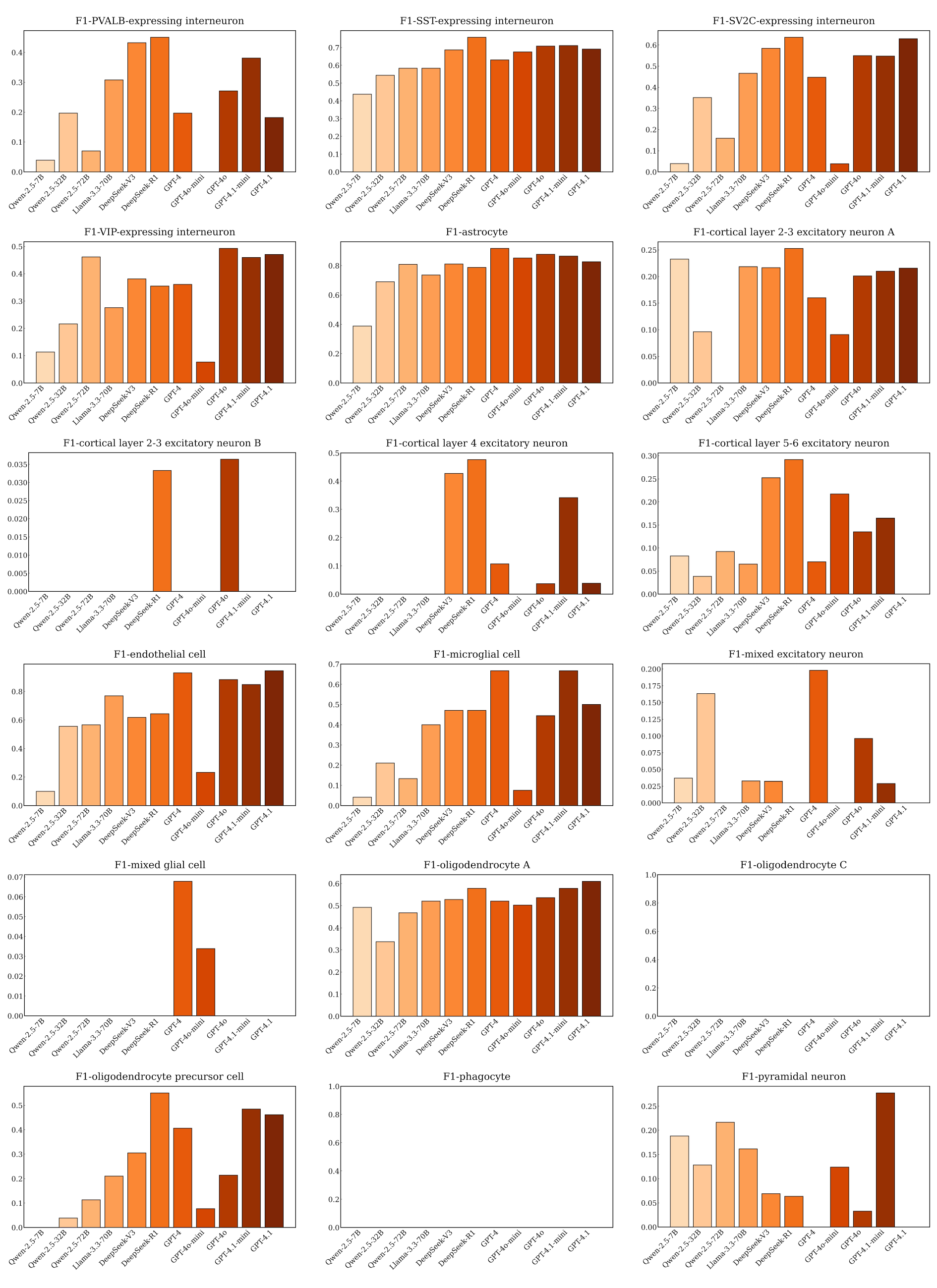}
    \caption{F1 score comparison of specific cell types on scRNA-seq data. All LLMs fail to predict certain challenging cell types, such as oligodendrocyte C and phagocyte.}
    \label{fig:f1_ms}
\end{figure}

\begin{figure}[h]
    \centering
    \includegraphics[width=\linewidth]{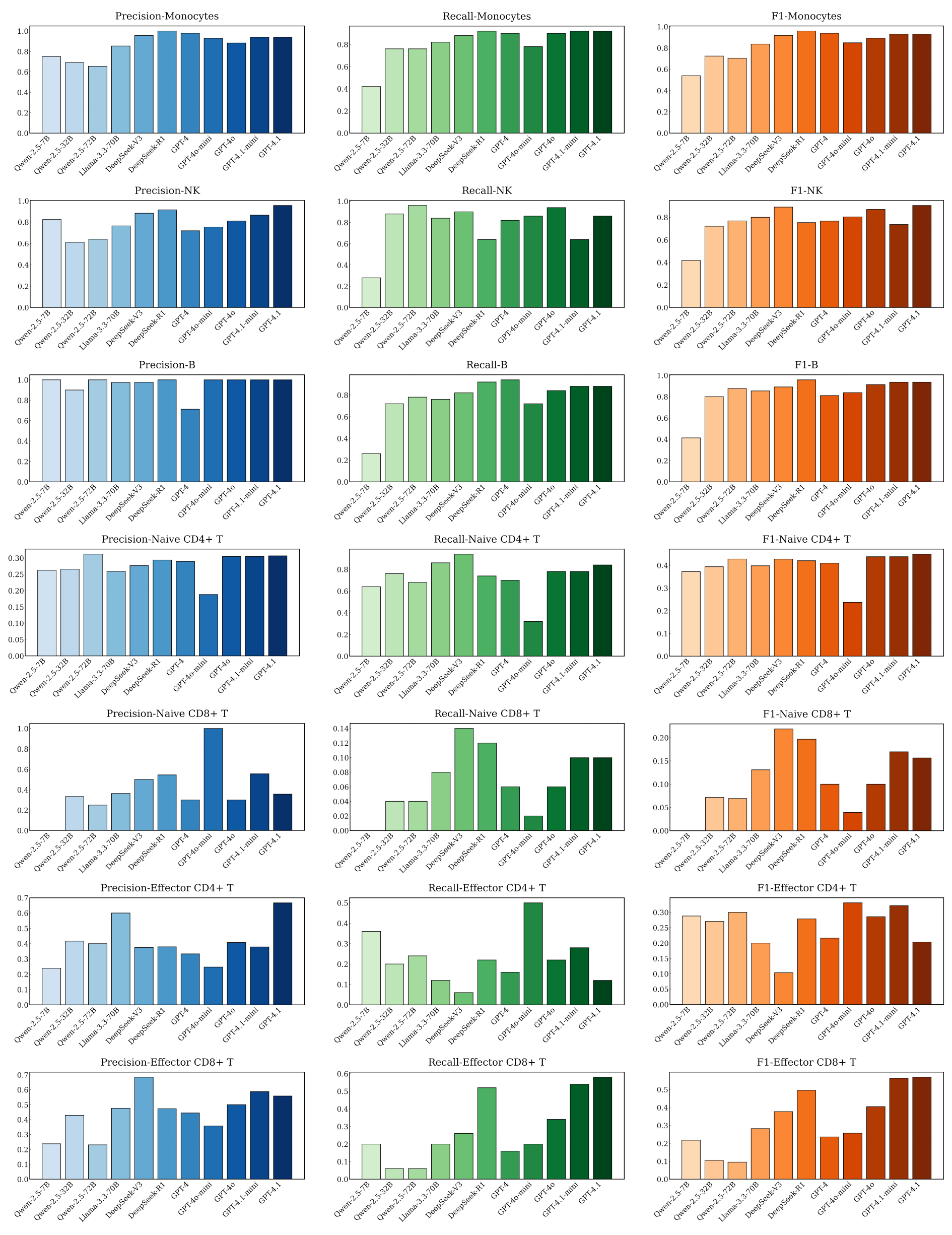}
    \caption{Performance comparison of specific cell types on CITE-seq data.}
    \label{fig:cite}
\end{figure}

\begin{figure}[h]
    \centering
    \includegraphics[width=\linewidth]{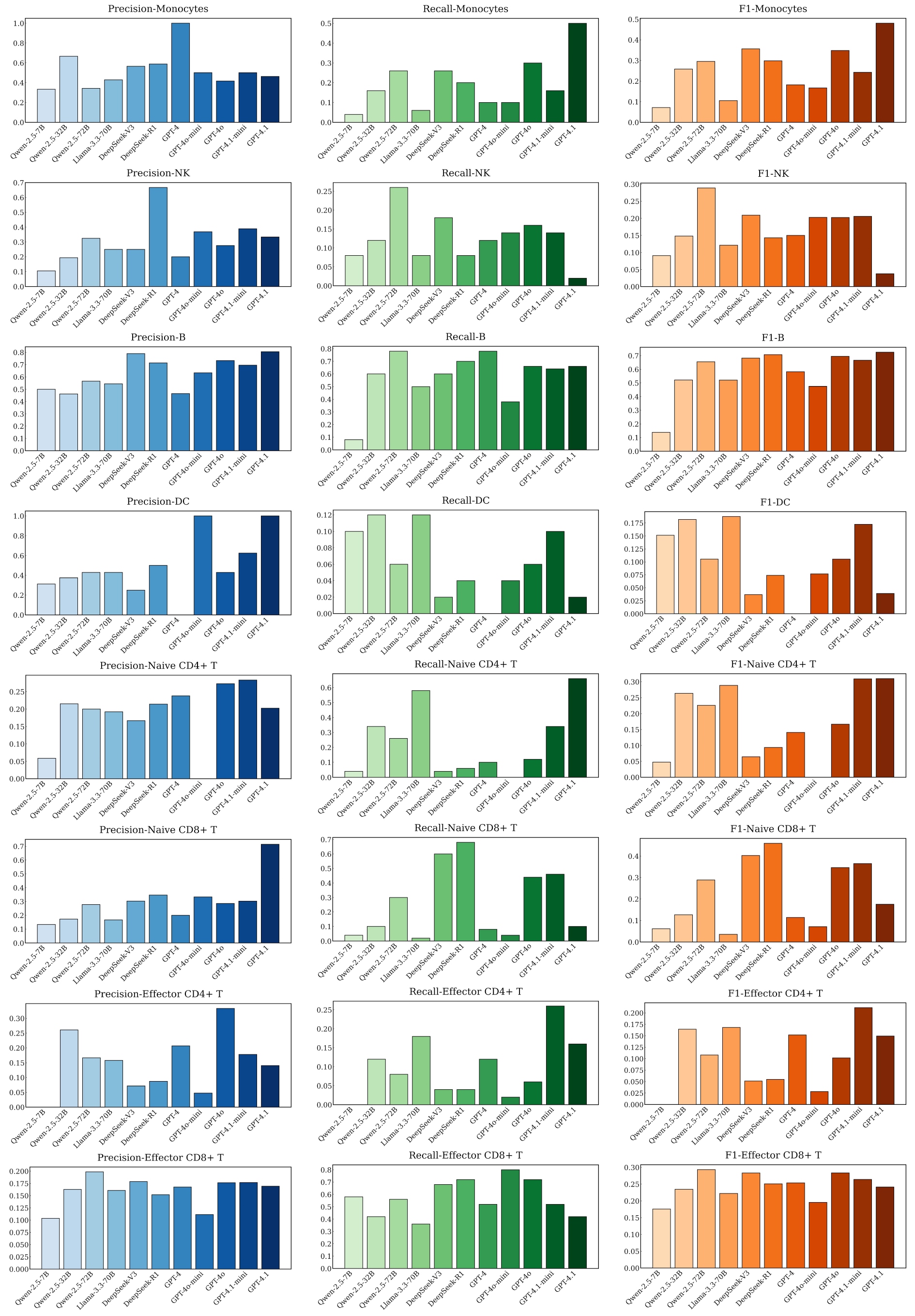}
    \caption{Performance comparison of specific cell types on ASAP-seq data.}
    \label{fig:asap}
\end{figure}

We present fine-grained comparisons of model performance across specific cell types in the cell type annotation task. Figures~\ref{fig:precision_ms}, \ref{fig:recall_ms}, and \ref{fig:f1_ms} report precision, recall, and F1 scores on scRNA-seq data, respectively. Similarly, Figures~\ref{fig:cite} and~\ref{fig:asap} show results on CITE-seq and ASAP-seq data. From the results, it can be observed that all the evaluated LLMs consistently struggle to identify certain challenging cell types, such as oligodendrocyte C and phagocyte.

\subsection{More Metrics Comparison under Few-shot Settings}
\begin{figure}[t]
    \centering
    \includegraphics[width=\linewidth]{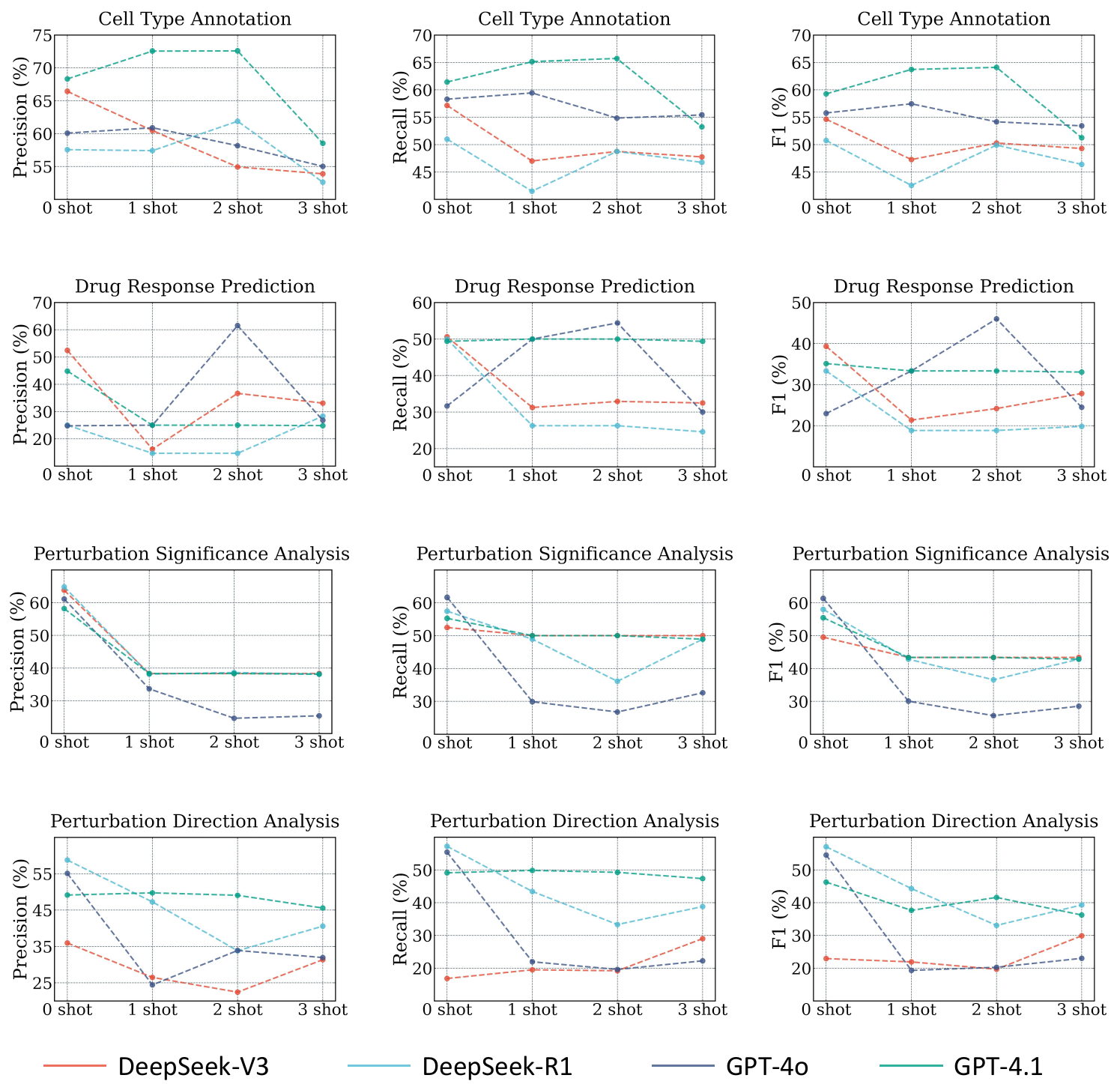}
    \caption{Precision, Recall, and F1 score comparison under few-shot settings.}
    \label{fig:few_shot_add}
\end{figure}

Figure \ref{fig:few_shot_add} presents experimental results for precision, recall, and F1 score under few-shot settings across different single-cell analysis tasks. The results indicate that for all of the evaluated LLMs, increasing the number of few-shot samples does not necessarily lead to performance gains and can even degrade model performance, which is consistent with our earlier analysis. These findings suggest that in future research on LLMs for cell biology, selecting high-quality and representative few-shot samples will be a critical and challenging problem.

\section{Additional Error Analysis}\label{sec:appendix_error_ana}
\begin{figure}[h]
    \centering
    \includegraphics[width=\linewidth]{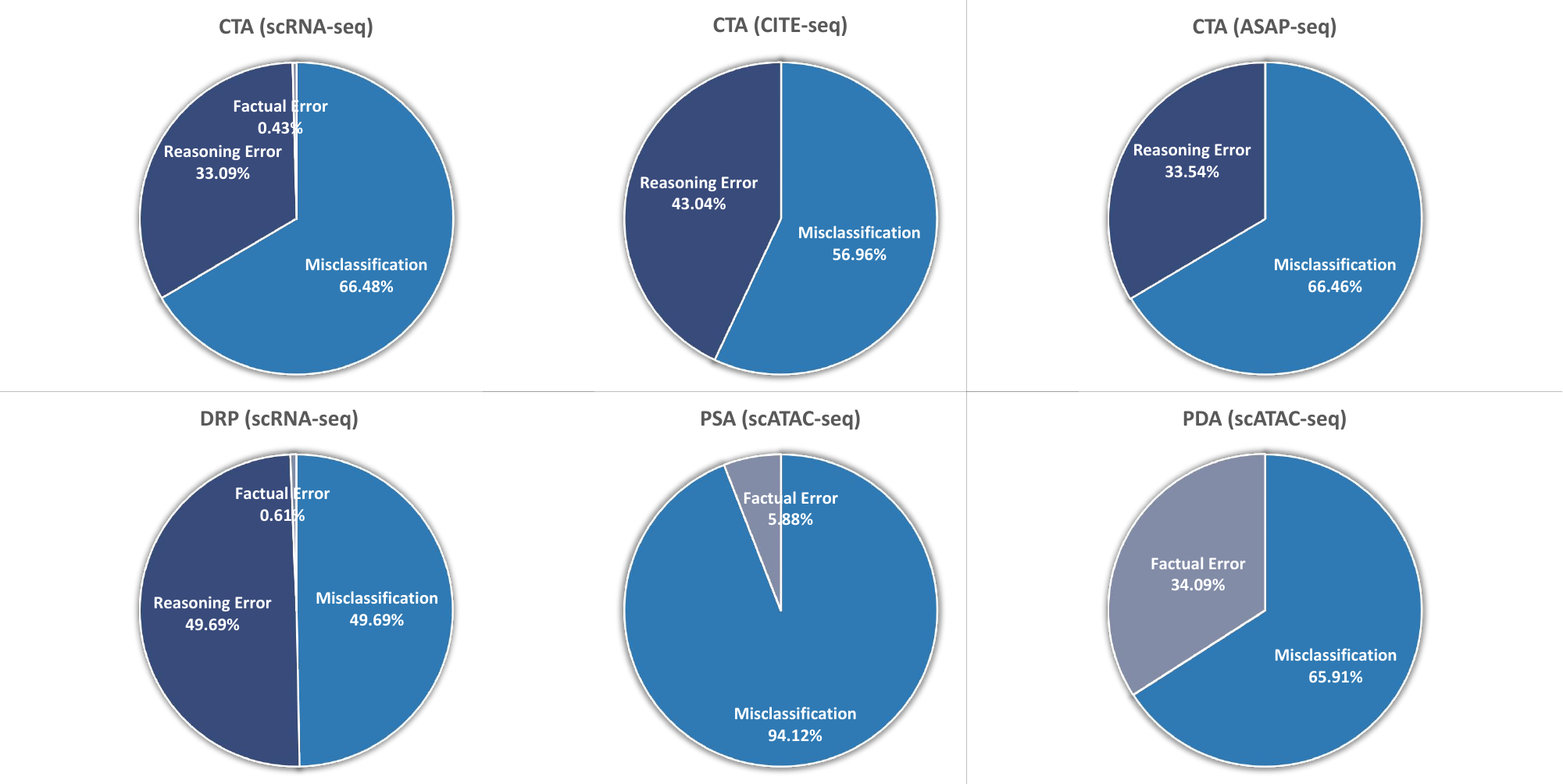}
    \caption{Distribution of \llmname{GPT-4.1}'s errors within distinct types across various tasks.}
    \label{fig:error_type_gpt41}
\end{figure}

\begin{figure}[h]
    \centering
    \includegraphics[width=\linewidth]{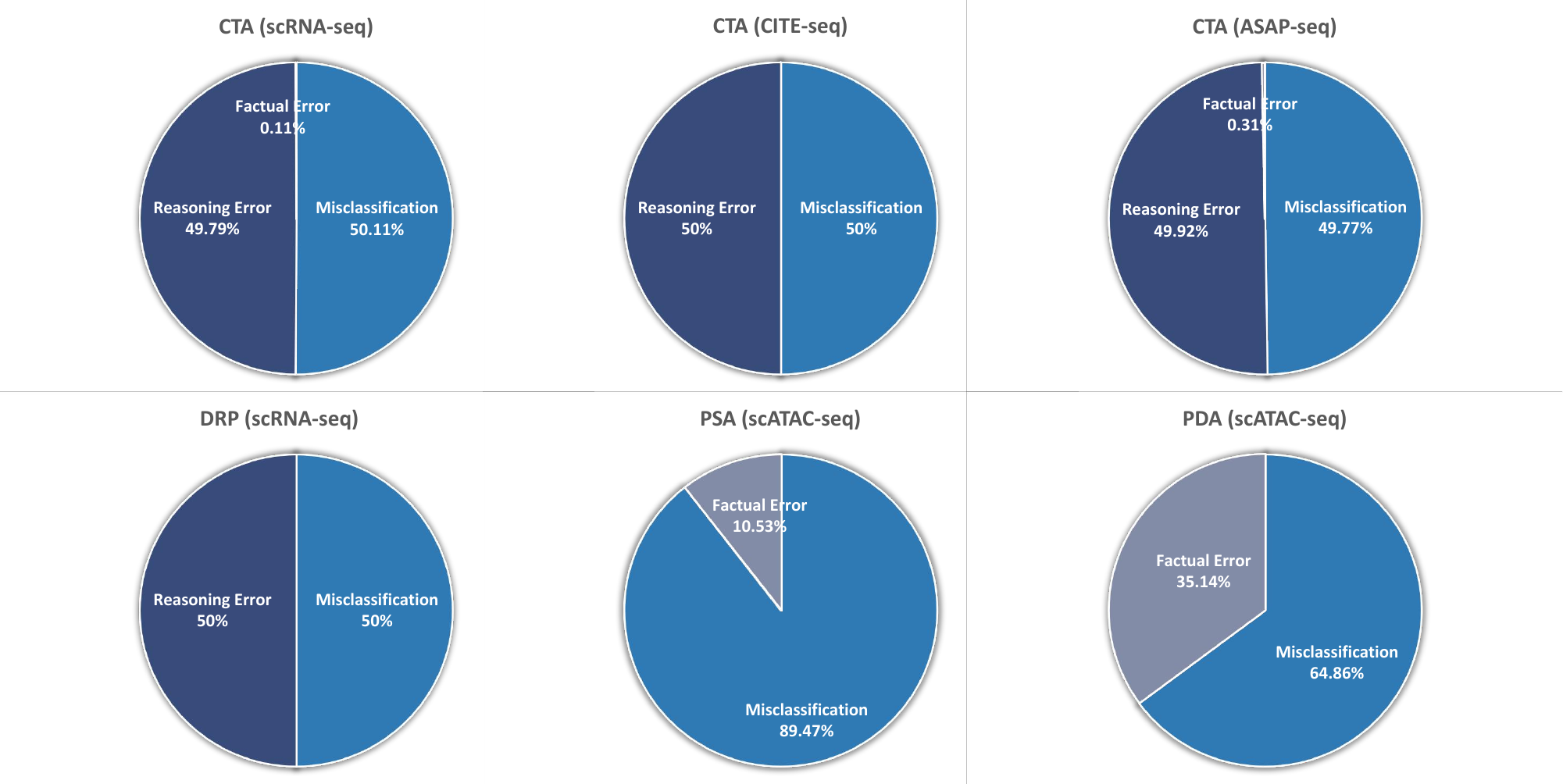}
    \caption{Distribution of \llmname{GPT-4o}'s errors within distinct types across various tasks.}
    \label{fig:error_type_gpt4o}
\end{figure}

In Figure \ref{fig:error_type_gpt41} and Figure \ref{fig:error_type_gpt4o}, we additionally present the error type distributions of \llmname{GPT-4.1} and \llmname{GPT-4o} across various tasks in \method{}. The results show that, similar to the error type distribution of \llmname{DeepSeek-R1}, misclassification and reasoning errors are also the two most common error types for \llmname{GPT-4.1} and \llmname{GPT-4o}. For cell-level tasks, factual errors are relatively rare, whereas their frequency increases in drug-level and gene-level tasks. This trend is expected, as drug-level and gene-level tasks are inherently more complex than cell-level ones.


\section{Case Study}\label{sec:appendix_case}
Here, we present a case study showcasing responses from different LLMs across various tasks in \method{}, demonstrating their reasoning capabilities in cell biology.
\begin{tcolorbox}[notitle, sharp corners, colframe=Periwinkle, colback=white, 
       boxrule=3pt, boxsep=0.5pt, enhanced, 
       shadow={3pt}{-3pt}{0pt}{opacity=1,mygrey},
       title={Cell Type Annotation},]\label{case:cta}
       \small
\textbf{Question:} Below is a list of \{gene\_names\} ordered by descending expression level in a multiple sclerosis cell. Your task is to select one cell type to which this cell most likely belongs from the following candidate cell types: \{cell\_types\}.\\
\{gene\_names\}: MT-CO2 ATP1B1 MT-CO3 MT-CO1 YWHAG RAB3B GAD1 MAP1B PSAP......\\
\{cell\_types\}: phagocyte, cortical layer 2-3 excitatory neuron A, cortical layer 4 excitatory neuron......\\
\textbf{Ground Truth:} PVALB-expressing interneuron\\

\llmname{DeepSeek-R1}:
The cell exhibits high expression of genes associated with GABAergic interneurons, including GAD1 and GAD2, which are critical for GABA synthesis. Synaptic markers like SNAP25, SYT1, and VAMP2 further confirm neuronal identity. Notably, KCNC2 (encoding Kv3.2, a potassium channel linked to fast-spiking interneurons) is present, which is characteristic of PVALB-expressing interneurons. While PVALB itself is not explicitly listed, the combination of GABAergic markers and KCNC2 strongly supports this subtype. Other interneuron subtypes (e.g., SST, VIP) lack their defining markers in the gene list. Glial or oligodendrocyte markers are absent, ruling out non-neuronal types.\\
\textbf{Final Answer:} PVALB-expressing interneuron \textcolor{green}{\ding{51}}\\

\llmname{GPT-4.1}:
Reasoning: The gene list is highly enriched for neuronal markers (e.g., SNAP25, SYT1, GAD1, GAD2, ERBB4, CALM3, ELAVL2, NDRG4, ATP1A3, KCNC2, DNER, PNMA2, ATP2A2, RAB3A, UCHL1, PRKACB, PRNP, RTN1, CPLX1, VAMP2, SLC24A2, STXBP1, GPM6A, SYN2, FGF12, DNM3, SLC6A1, SLC6A17, SYN1, SCAMP5, GABRA1, GABRG1, GLRA2, GLRA3, GRIK1, GRIK3, GABARAPL1, GABARAPL2, etc.). There is strong expression of GAD1 and GAD2, which are canonical markers of GABAergic interneurons. SV2C is not directly listed, but the presence of many synaptic vesicle and GABAergic genes (e.g., RAB3A/B/C, SNAP25, SYT1, STXBP1, VAMP2, SLC6A1, SLC6A17) supports an interneuron identity. There is no strong evidence for astrocyte, oligodendrocyte, microglial, or endothelial markers. Among interneuron subtypes, there is no strong expression of VIP, SST, or PVALB, which would suggest those specific subtypes. The overall profile, with high GABAergic and synaptic gene expression, best matches the SV2C-expressing interneuron among the provided options.\\
\textbf{Final Answer:} SV2C-expressing interneuron \textcolor{red}{\ding{55}} \\

\llmname{GPT-4o}:
To determine the most likely cell type based on the gene expression profile provided, we need to identify key marker genes that are characteristic of specific cell types. 1. Neuronal Markers: Genes such as SNAP25, SYT1, MAP1B, and GAD1 are typically associated with neurons. GAD1 and GAD2 are markers for GABAergic neurons, which include interneurons like SST-expressing, VIP-expressing, and PVALB-expressing interneurons. 2. Excitatory Neuron Markers: Genes like ATP1A3, CAMK2G, and MAPK1 are often found in excitatory neurons, including cortical layer neurons and pyramidal neurons. 3. Interneuron Markers: SST, VIP, and PVALB are specific markers for different types of interneurons. 4. Glial Cell Markers: Genes such as GFAP, S100B, and AQP4 are typical for astrocytes. Oligodendrocyte markers include MBP, MOG, and PLP1, but none of these are prominently listed. Microglial markers include CX3CR1 and P2RY12, which are not prominently listed. 5. Other Cell Types: Endothelial cells often express genes like PECAM1 and VWF, which are not prominently listed. Given the presence of neuronal markers such as SNAP25, SYT1, and GAD1, the cell is likely a neuron. The presence of GAD1 and GAD2 suggests it is a GABAergic neuron, which points towards an interneuron. Among the interneurons listed, SST-expressing interneurons are a possibility due to the presence of genes like SST and other neuronal markers.\\
\textbf{Final Answer:} SST-expressing interneuron \textcolor{red}{\ding{55}} 
\end{tcolorbox}

\begin{tcolorbox}[notitle, sharp corners, colframe=ForestGreen, colback=white, 
       boxrule=3pt, boxsep=0.5pt, enhanced, 
       shadow={3pt}{-3pt}{0pt}{opacity=1,mygrey},
       title={Drug Response Prediction},]\label{case:drp}
       \small
\textbf{Question:} Below is a list of \{gene\_names\} ordered by descending expression level in a cell. Your task is to predict its response to the drug Erlotinib: [Resistant, Sensitive].\\
\{gene\_names\}: HSP90AA4P MALSU1 RPL37A FTLP1 PABPC1L RPLP1P13 RPS16P5 RPS14P3......\\
\textbf{Ground Truth:} Sensitive\\

\llmname{DeepSeek-R1}:
The prediction of Erlotinib response hinges on identifying markers linked to resistance or sensitivity. Erlotinib targets EGFR, with sensitivity often tied to activating EGFR mutations (e.g., exon 19 deletions, L858R) and resistance to mechanisms like T790M mutations, MET amplification, or bypass signaling pathways. The provided gene list lacks EGFR itself but includes potential resistance-associated genes: 1. RPS6KA1 (Ribosomal S6 Kinase A1): This kinase is part of the MAPK/ERK pathway downstream of EGFR. High expression suggests sustained ERK signaling independent of EGFR inhibition, a known resistance mechanism. Persistent MAPK pathway activity could bypass EGFR blockade, promoting survival. 2. HSP90 Pseudogenes (HSP90AA4P, HSP90AB2P, HSP90B2P): HSP90 stabilizes oncoproteins, including mutant EGFR. Elevated HSP90 expression might stabilize resistance-associated EGFR variants (e.g., T790M) or other client proteins, enabling survival despite EGFR inhibition. 3. S100A9/S100A11: These genes are implicated in inflammation and resistance pathways, potentially modulating survival signals or drug efflux, though their direct role here is unclear. 4. Ribosomal and Mitochondrial Genes (RPL, RPS, MT-CO3, MT-CYB): High ribosomal activity may indicate proliferation, while mitochondrial genes suggest metabolic adaptation, both common in resistant phenotypes. However, these are broadly expressed and less specific. The absence of clear sensitivity markers (e.g., EGFR mutations, epithelial markers) and the presence of RPS6KA1 (sustaining MAPK signaling) and HSP90 pseudogenes (stabilizing oncoproteins) strongly suggest resistance. While ribosomal and mitochondrial genes are non-specific, their prominence aligns with proliferative and adaptive states typical of resistant cells.\\
\textbf{Final Answer:} Resistant \textcolor{red}{\ding{55}}\\

\llmname{GPT-4.1}:
To predict the response to Erlotinib, we need to consider the expression of genes associated with sensitivity or resistance to this drug. Erlotinib is an EGFR (Epidermal Growth Factor Receptor) inhibitor, and sensitivity is often associated with high EGFR pathway activity and low expression of resistance markers such as EMT (epithelial-mesenchymal transition) genes, AXL, MET, or high expression of certain ribosomal and stress response genes. From the provided gene list, the most highly expressed genes are ribosomal proteins (RPL, RPS), heat shock proteins (HSP90AA4P, HSP90AB2P, HSP90B2P), and other housekeeping genes. There is no evidence of high expression of EGFR, AXL, MET, or EMT markers (such as VIM, ZEB1, SNAI1, TWIST1). However, high expression of ribosomal proteins and heat shock proteins can sometimes be associated with a more proliferative, less differentiated state, which may be more sensitive to EGFR inhibition, but high HSP90 expression can also be associated with resistance mechanisms. Notably, there is no clear evidence of canonical resistance markers (e.g., AXL, MET, EMT genes) or sensitivity markers (e.g., high EGFR, ERBB2, or downstream effectors). Given the lack of strong resistance markers and the predominance of ribosomal and housekeeping genes, the cell is more likely to be in a proliferative state, which is generally more sensitive to Erlotinib.\\
\textbf{Final Answer:} Sensitive \textcolor{green}{\ding{51}}\\

\llmname{GPT-4o}:
To predict the response of a cell to the drug Erlotinib based on gene expression data, we need to consider the known mechanisms of action of Erlotinib and the role of the genes listed. Erlotinib is an EGFR (epidermal growth factor receptor) inhibitor, and its effectiveness can be influenced by the expression of genes involved in the EGFR signaling pathway, as well as genes related to cell proliferation, apoptosis, and drug resistance mechanisms. In the provided list, there are several ribosomal protein genes (e.g., RPL, RPS), heat shock proteins (e.g., HSP90), and other genes involved in cellular stress responses and protein synthesis. High expression of ribosomal proteins and heat shock proteins can sometimes be associated with increased cellular proliferation and survival, which might contribute to drug resistance. However, without specific information on the expression of genes directly involved in the EGFR pathway or known resistance markers, it is challenging to make a definitive prediction. Generally, high expression of genes that promote cell survival and proliferation could suggest a resistant phenotype, but this is a complex decision that would ideally be informed by more specific data. Given the high expression of heat shock proteins and ribosomal proteins, which can be associated with increased cellular resilience and potential drug resistance, I would lean towards predicting resistance.\\
\textbf{Final Answer:} Resistant \textcolor{red}{\ding{55}} 
\end{tcolorbox}

\begin{tcolorbox}[notitle, sharp corners, colframe=YellowGreen, colback=white, 
       boxrule=3pt, boxsep=0.5pt, enhanced, 
       shadow={3pt}{-3pt}{0pt}{opacity=1,mygrey},
       title={Perturbation Significance Analysis},]\label{case:psa}
       \small
\textbf{Question:} Does perturbing NEDD8 in K562 cells cause significant changes in the expression of GFI1B?\\
\textbf{Ground Truth:} Yes\\

\llmname{DeepSeek-R1}:
To determine if perturbing NEDD8 in K562 cells significantly alters GFI1B expression, we consider the role of NEDD8 in neddylation, a process critical for cullin-RING ubiquitin ligase (CRL) activity. CRLs mediate protein degradation, and their inhibition via NEDD8 perturbation can stabilize or destabilize transcription factors regulating GFI1B. Experimental evidence, such as studies using the neddylation inhibitor MLN4924 in K562 cells, has shown that GFI1B expression is downregulated following NEDD8 inhibition. This suggests that NEDD8 perturbation disrupts CRL-mediated degradation pathways, indirectly affecting GFI1B levels.\\
\textbf{Final Answer:} Yes \textcolor{green}{\ding{51}}\\

\llmname{GPT-4.1}:
No. \\
\textbf{Final Answer:} No \textcolor{red}{\ding{55}}\\

\llmname{GPT-4o}:
Yes. \\
\textbf{Final Answer:} Yes \textcolor{green}{\ding{51}} 
\end{tcolorbox}

\begin{tcolorbox}[notitle, sharp corners, colframe=Melon, colback=white, 
       boxrule=3pt, boxsep=0.5pt, enhanced, 
       shadow={3pt}{-3pt}{0pt}{opacity=1,mygrey},
       title={Perturbation Direction Analysis},]\label{case:pda}
       \small
\textbf{Question:} Does perturbing QARS in K562 cells cause the expression of PCM1 to upregulate or downregulate?\\
\textbf{Ground Truth:} Up\\

\llmname{DeepSeek-R1}:
To determine whether perturbing QARS in K562 cells affects PCM1 expression, we consider the roles of these genes and potential pathways. QARS, a glutaminyl-tRNA synthetase, is critical for protein synthesis, and its perturbation could activate stress responses like the integrated stress response (ISR) via GCN2 kinase. This might suppress general translation but upregulate stress-related genes. PCM1, involved in centriole assembly, could be influenced by cell cycle changes or structural stress. While direct evidence is lacking, amino acid deprivation stress (via QARS inhibition) often downregulates proliferation-related genes. PCM1's role in cell division suggests potential downregulation. However, stress responses might also upregulate structural genes as compensatory mechanisms. Considering studies on tRNA synthetase inhibition leading to stress-induced gene upregulation, PCM1 might be upregulated. After evaluating these pathways and potential compensatory mechanisms, the balance leans toward upregulation under stress.\\
\textbf{Final Answer:} Up \textcolor{green}{\ding{51}}\\

\llmname{GPT-4.1}:
Down. \\
\textbf{Final Answer:} Down \textcolor{red}{\ding{55}}\\

\llmname{GPT-4o}:
Down. \\
\textbf{Final Answer:} Down \textcolor{red}{\ding{55}} 
\end{tcolorbox}

\section{Limitation and Future Work}\label{sec:appendix_limit}
Although \method{} is the first language-centric benchmark for single-cell analysis with LLMs and marks a step forward in applying LLMs to cell biology, it still presents several noteworthy limitations.

First, while \method{} spans hierarchical single-cell analysis tasks, such as cell type annotation (cell-level), drug response prediction (drug-level), and perturbation analysis (gene-level), it does not yet provide a quantitative distinction in their levels of difficulty. Future work could draw inspiration from benchmarks in mathematics \cite{hendrycks2021measuring} to assign difficulty levels to distinct problems, enabling a more nuanced understanding of the problem-solving capabilities of LLMs in cell biology.

Second, although \method{} introduces a pipeline for converting raw single-cell multi-omics data into natural language QA problems across multiple tasks, all prompts and questions are currently formulated in English. Extending the benchmark to include multilingual QA settings would improve its global applicability and allow for more comprehensive evaluation of LLMs in terms of linguistic diversity and understanding, which we leave as future work.

\end{document}